\documentclass[10pt]{iopart}
\usepackage{graphicx}
\usepackage{subfigure}
\graphicspath{{Figures/}}
\usepackage{amsmath}
\usepackage{amssymb}
\usepackage{hyperref}
\usepackage{bm}
\usepackage[dvipsnames]{xcolor}
\def\eg{{\it e.g., }}
\def\ie{{\it i.e., }}
\newcommand{\etc}{{etc. }}              % etc.
\everymath{\displaystyle}

\DeclareMathOperator*{\Argmin}{Argmin}

\newcommand{\revision}[1]{{#1}}

\begin{document}
\title[]{Deformable registration with intensity correction for CESM monitoring response to Neoadjuvant Chemotherapy}
\author{
Clément Jailin$^1$, Pablo Milioni De Carvalho$^1$, 
Sara Mohamed$^1$, Laurence Vancamberg$^1$,
Amr Farouk Ibrahim Moustafa$^2$,
Mohammed Mohammed Gomaa$^2$,
Rasha Mohammed Kamal$^2$, Serge Muller$^1$}

\address{  $^1$ GE Healthcare, 78530 Buc, France\\
  $^2$ Baheya Foundation For Early Detection And Treatment Of Breast Cancer, El Haram, Giza,  Egypt\\
}
\ead{clement.jailin@ge.com}

\begin{abstract}
This paper proposes a robust longitudinal registration method for Contrast Enhanced Spectral Mammography in monitoring neoadjuvant chemotherapy. Because breast texture intensity changes with the treatment, a non-rigid registration procedure with local intensity compensations is developed. The approach allows registering the low energy images of the exams acquired before and after the chemotherapy. The measured motion is then applied to the corresponding recombined images. 
The difference of registered images, called residual, makes vanishing the breast texture that did not changed between the two exams. 
Consequently, this registered residual allows identifying local density and iodine changes, especially in the lesion area.
The method is validated with a synthetic NAC case where ground truths are available. Then the procedure is applied to 51 patients with 208 CESM image pairs acquired before and after the chemotherapy treatment. The proposed registration converged in all 208 cases.  
The intensity-compensated registration approach is evaluated with different mathematical metrics and through the repositioning of clinical landmarks (RMSE: 5.9~mm) and outperforms state-of-the-art registration techniques.
\end{abstract}

% ------------------
\vspace{1pc}

% ------------------------------------------------------------------------------
%
%   INTRODUCTION
%
% ------------------------------------------------------------------------------

\section{Introduction}

% ------------------
Neoadjuvant chemotherapy (NAC) is a therapeutic option increasingly used in the management strategy for breast cancer. 
Imaging the breast before, (during) and after the chemotherapy is important to evaluate the treatment response and future treatment planning. Lesion evolution information (\eg lengths, extent, intensity) can be measured and discussed with the oncologist and surgeon during the tumor board.

% ------------------
Magnetic resonance imaging, MRI (and contrast-enhanced MRI) is an excellent imaging modality to monitor response to neoadjuvant chemotherapy~\cite{fowler2017imaging}. Various studies have been developed to extract quantitative features in pre-NAC and/or post-NAC images related to lesion evolution~\cite{ou2015deformable,liu2019radiomics, jahani2019prediction} and to identify breast tumor bed location for adjuvant therapy. 

% ------------------
Because pre-NAC and post-NAC images are not acquired in the same position, it is a beneficial task to register images for easier texture-to-texture comparison.
Registration in (breast) NAC-MRI has been developed in the literature and is often based on optical flow~\cite{ou2015deformable}. Registered images can be used for quantitative assessment. The intensity of the registered textures can be compared pixel-wised~\cite{jahani2019prediction,fan2021radiomics} and the identified motion (or its Jacobian~\cite{riyahi2018quantifying,riyahi2018quantification}) may be related to the dynamic of the lesion shape~\cite{wodzinski2018improving}.

%However, it is difficult to quantitatively exploit both the identified motion and the registered MRI images. 
The lesion size change can be read as motion and/or intensity evolution. 
The difficulty comes when the measured motion corrects part of the lesion shape changes without succeeding in perfectly registering it. The lesion evolution information is hence split both in an imperfect measured motion field and in a partially registered image. This split is not exploitable and makes the analysis more qualitative than quantitative.

% ------------------
Contrast-Enhanced Spectral Mammography (CESM) provides anatomical and functional imaging of breast tissue improving the accuracy of breast cancer diagnosis~\cite{dromain2011dual}. 
The competence of CESM in the assessment and prediction of response to NAC has been recently studied~\cite{iotti2017contrast,barra2018contrast,patel2018contrast,steinhof2021contrast} using image lesion measurements (\eg RECIST~\cite{eisenhauer2009new}).
In addition to the measurement of lesion lengths, some studies presented a method extracting quantitative features from recombined CESM images to evaluate residual disease extents (lesion characteristics with intensity measurements in~\cite{moustafa2019quantitative,kamal2020predicting} and radiomics features in~\cite{xing2021quantitative}). In the study performed by Wang \textit{et.al}~\cite{wang2021contrast}~, the authors aimed to predict from the radiomics extraction on the pre-NAC exam image the chemotherapy results. Because  CESM images were not textured aligned, the intensity comparison can just be global in the annotated region of interest. Local texture changes could not be easily captured.

Developing a registration procedure was proved to be efficient in MRI for quantitative features extraction and for identifying breast tumor bed location.  
Registering similar textures in CESM images will help clinicians in identifying texture changes and will improve NAC quantification. 
When images are registered, the pixel intensity differences can be read as texture differences. The major part of the texture which is similar to the two registered images vanishes. The local texture changes are hence highlighted.

% ------------------
Registration in 2D mammography has been widely studied for decades in the literature. The soft nature of breast tissue requires the registration to be non-rigid~\cite{van2003comparison}. 
%Often based on 2D corrections of full field digital mammograms (FFDM), it may also be extended to 3D (\eg digital breast tomosynthesis - DBT, computed tomography scans - CT, magnetic resonance imaging - MRI, ultrasounds, \etc). 
Many registration approaches have been developed such as feature-based~\cite{li2015bilateral}, intensity-based~\cite{richard2003new,diez2011revisiting,garcia2017similarity} and deep-learning-based~\cite{zhang2019mammographic,haskins2020deep} methods.
%(although with mammogram registration, sub-categories may be seen: \emph{contour-based}~\cite{marias2005registration} and \emph{anatomical structures-based} approaches).
When registered, the image difference may be related to the time growth of a density or the apparition of micro-calcifications~\cite{marti2014detecting,mendel2019temporal,robinson2019machine}. 
%In~\cite{9035423}, for example, the authors used the segmented registered longitudinal differences to highlight the apparition of micro-calcifications.

% ------------------
The major limitations of state-of-the-art breast registration approaches are (i) large displacements and (ii) important intensity variations. 
Those issues are exacerbated when dealing with NAC cases composed of large tumors that may shrink or disappear (for complete pathological responses). This bias affects the registration and limits its quantitative use.

% ------------------
The goal of this paper is to propose a methodology to register NAC-CESM data  (i) with large repositioning displacements, (ii) with important intensity changes, and (iii) with a high convergence rate to be potentially used in a clinical environment.
The first section presents the NAC-CESM database. Then, different registration methods are described: a state-of-the-art TV-L1 algorithm, the global digital image correlation (GDIC) method, and the proposed registration method (GDIC-I) enhanced with local intensity corrections. 
%After validation on synthetic data, the procedure was evaluated on 51 patients who underwent chemotherapy (corresponding to 208 CESM image pairs) based on different mathematical and clinical metrics. 
%The GDIC-I procedure that converged on all NAC cases outperforms standard registration approaches.

% ------------------------------------------------------------------------------
%
%   PART 1 : DATA
%
% ------------------------------------------------------------------------------

\section{Materials}

% --------------------------
\subsection{NAC clinical practice}
The main goal of NAC is the reduction of tumor volume and metastasis leading to an increase in breast-conserving surgery probabilities instead of mastectomy. NAC also allows early assessment of the efficiency of systemic therapy in-vivo which could lead to a revision of the treatment plan in cases that show poor response. The extent of residual invasive cancer after neoadjuvant therapy is also a prognostic factor for the risk of recurrence.

In CESM-NAC monitoring, the patient is imaged with a CESM acquisition pre-NAC and post-NAC.
The CESM acquisition consists of a low-energy (LE) image and a high-energy (HE) image processed into a so-called recombined image that highlights the iodine content. In our study, the image size is 2394$\times$2850~pixels with a 100~$\mu$mm pixel pitch. 

After the acquisition, lesion size and extent are reported in both pre-NAC and post-NAC studies. Radiological response to NAC is assessed once applying the Response Evaluation Criteria in Solid Tumor~\cite{eisenhauer2009new}: RECIST 1.1. This evaluation is using the ratio of the longest length of the lesion before and after NAC. The ratio is read in 4 classes: complete response (pCR), partial response (more than 30\% decrease), stable disease (up to 20\% increase), and progressive disease.
Being able to evaluate the lesion evolution is important for patient treatment planning.

\subsection{Synthetic NAC case}
A test case is generated to be realistic and to provide a known ground truth of a partial NAC radiological response (RECIST: 50\% decrease). 
For this test case, a LE (or FFDM) acquisition $f_0(\bm x)$ is used as breast texture, defined on all pixels $\bm x$. To create the pre-NAC image $f(\bm x)$, an synthetic pre-NAC lesion $\mathcal{L}_f(\bm x)$ is inserted: $f(\bm x)=f_0(\bm x)+\mathcal{L}_f(\bm x)$. 

\begin{figure}[t!]
	\centering
        \subfigure[]{\includegraphics[width=0.24\textwidth]{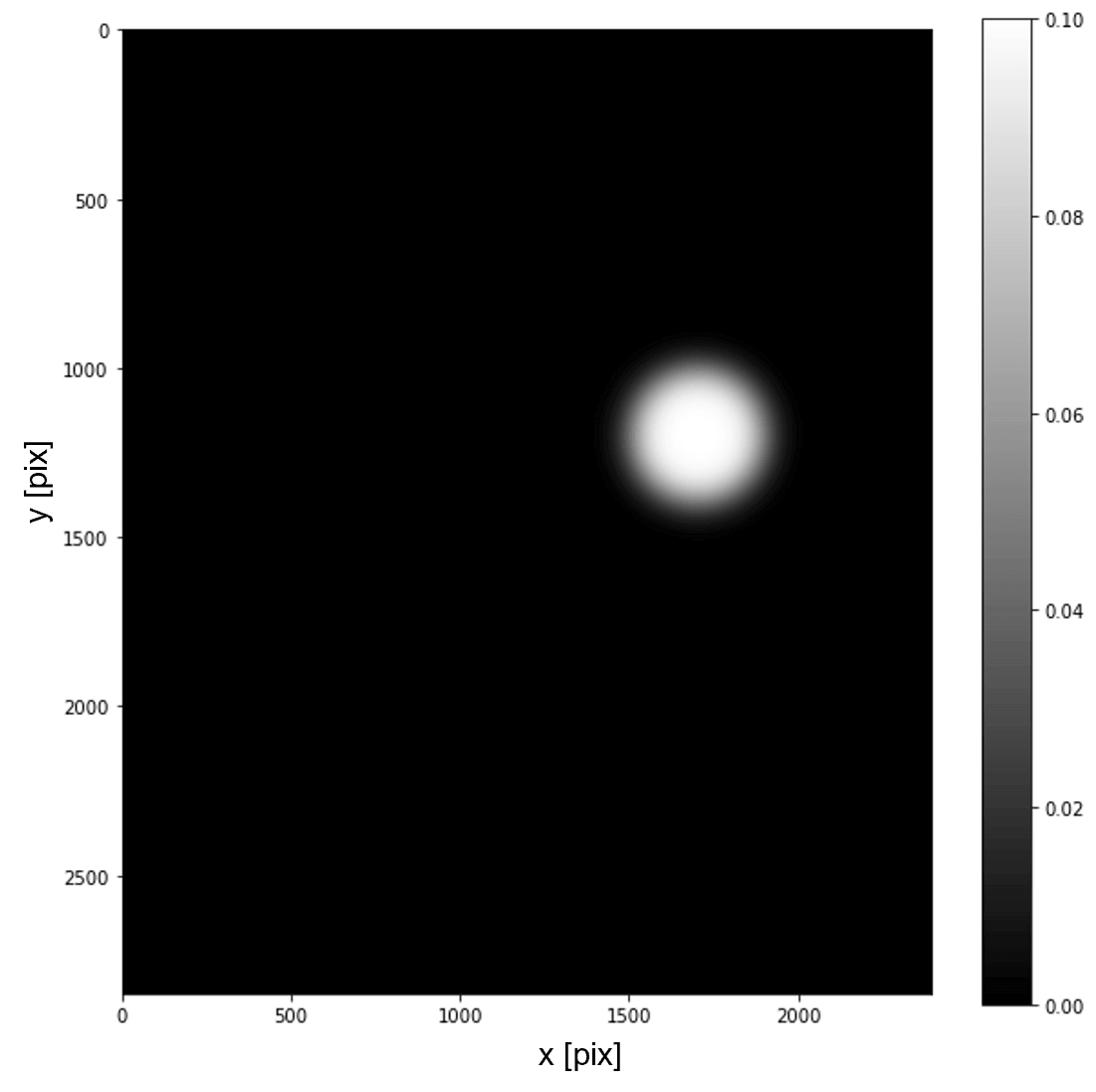}}
        \subfigure[]{\includegraphics[width=0.24\textwidth]{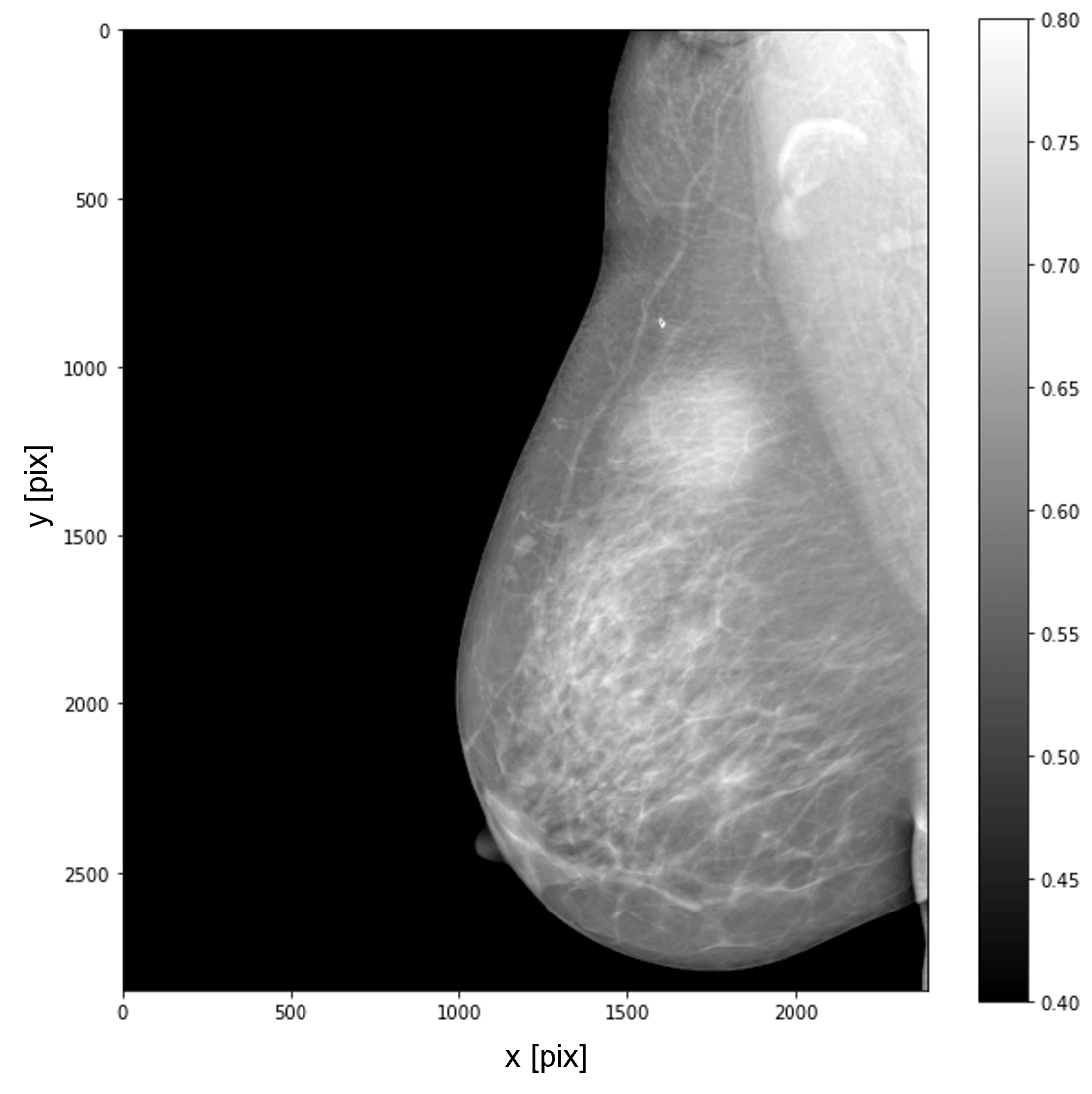}}
        \subfigure[]{\includegraphics[width=0.24\textwidth]{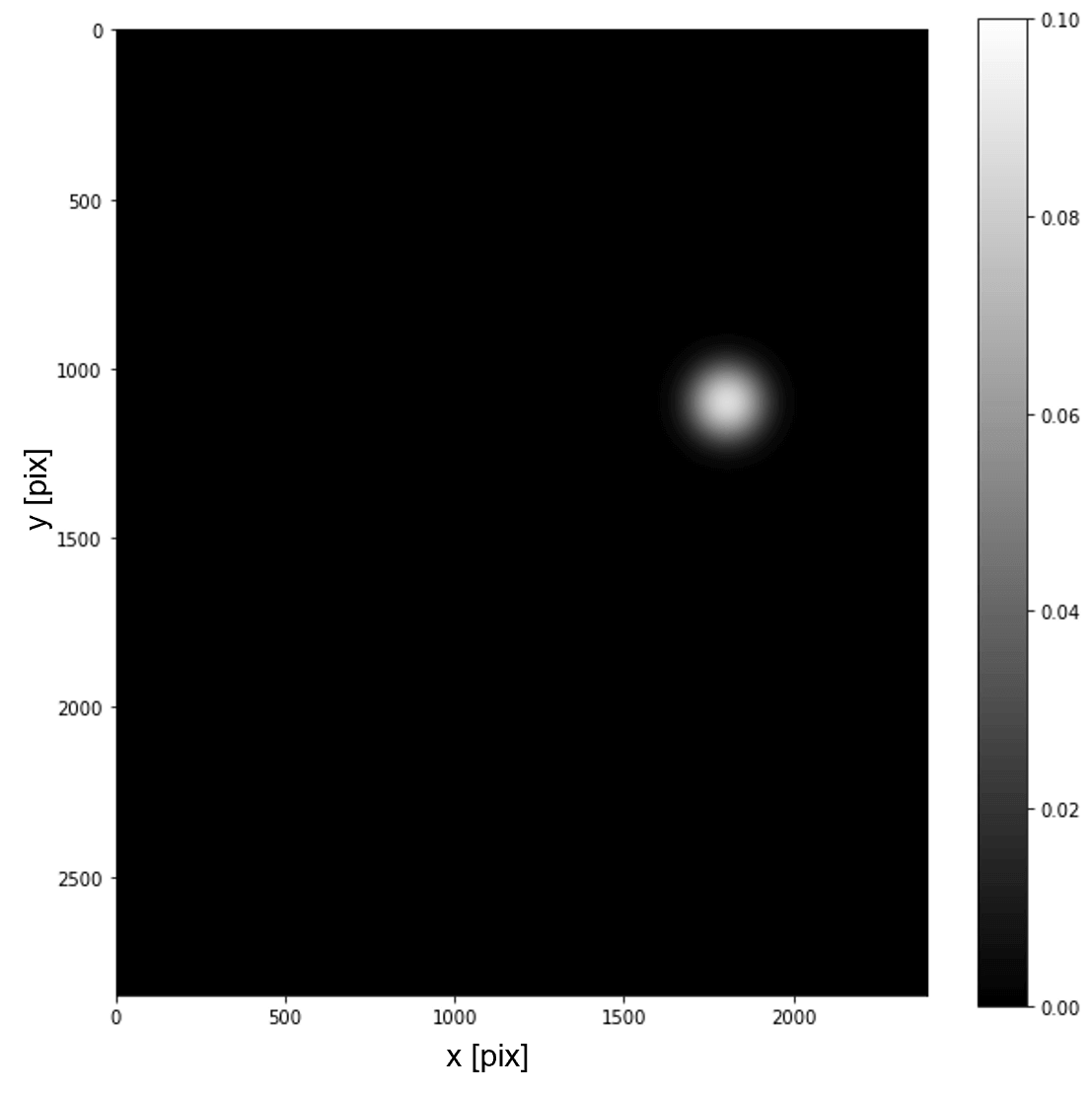}}
        \subfigure[]{\includegraphics[width=0.24\textwidth]{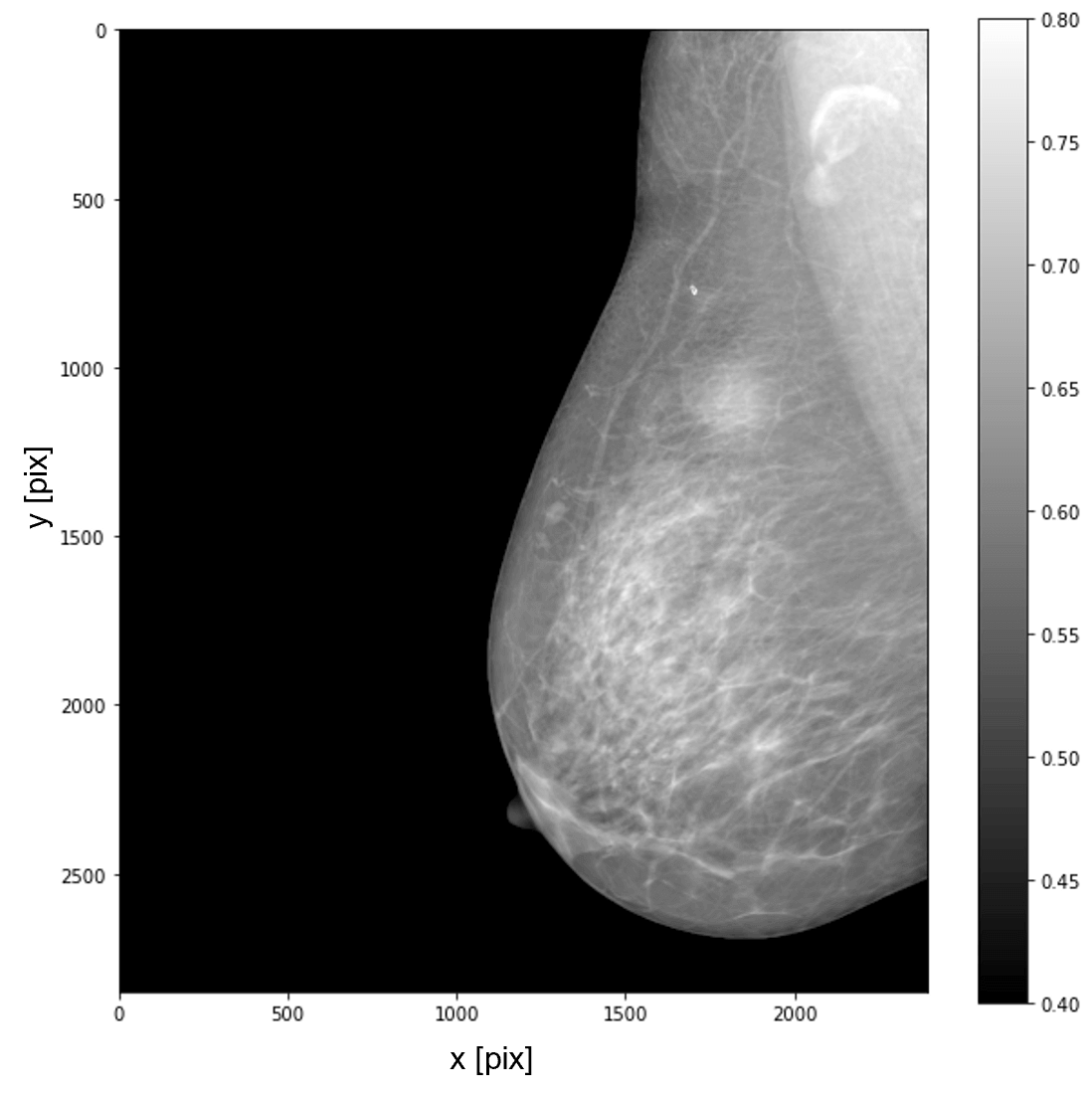}}
	\caption{Images used in the synthetic test case: (a)  pre-NAC lesion $\mathcal{L}_f(\bm x)$, (b) pre-NAC image $f(\bm x)=f_0(\bm x)+\mathcal{L}_f(\bm x)$, (c)  post-NAC lesion $\mathcal{L}_g(\bm x+ \bm u_{\text{gt}})$, and (c) post-NAC image $g(\bm x)=f_0(\bm x + \bm u_{\text{gt}})+\mathcal{L}_g(\bm x+ \bm u_{\text{gt}})$}
	\label{fig:synthetic1} 
\end{figure}

The post-NAC image is generated by adding to the breast texture image a post-NAC lesion $\mathcal{L}_g(\bm x)$ and by repositioning the image by a known 2D plane motion $u_{\text{gt}}$ such that $g(\bm x)=f_0(\bm x+ \bm u_{\text{gt}})+\mathcal{L}_g(\bm x+ \bm u_{\text{gt}})$.
All images are shown Figure~\ref{fig:synthetic1}.
For the test case, the motion field is a uniform translation of 1~cm (100~pixels) in both $x$ and $y$ spatial directions.

The lesion pre-NAC is a disk of 4~cm diameter, with a amplitude of 10\% of the gray level dynamic, convoluted by a Gaussian blur with a characteristic length of 4~mm (40 pixels). The lesion post-NAC is a disk of 2~cm diameter, with a amplitude of 10\% of the gray level dynamic, convoluted by a Gaussian blur with a characteristic length of 4~mm (40 pixels).
In this synthetic case, the simulated lesion has shrunk without change in the bulk intensity. The RECIST would be measured to 50\%. 

% --------------------------
\subsection{CESM-NAC cohort}

Fifty-one patients with pathologically proven breast cancer based on the tumor tissues obtained by core needle biopsy were enrolled in this retrospective study from March 2020 to March 2022. All data were acquired at \emph{Baheya Foundation For Early Detection And Treatment Of Breast Cancer, Giza, Egypt}.
They were all planned to receive NAC in reference to the breast cancer tumor board decision. All patients underwent two separate CESM examinations. The first examination was before beginning the NAC course and the second examination was after completing the NAC course. As both breasts were imaged with multiple views,  208 acquired image pairs were acquired.
The study protocol was approved by the Institutional Review Board and informed written consent was applied for the use data of the enrolled individuals.
The maximum interval between the post-NAC study and surgery was 10 days. All 52 biopsied lesions were invasive duct carcinoma.
Patients who were not candidates for NAC, patients with metastatic disease, pregnant females, and those who gave a history of allergy to contrast media, or renal impairment were omitted from the study.

Examinations were performed with a Senographe Pristina\texttrademark ~(\emph{GE Healthcare, Chicago, IL, USA}). The recombination was performed using the latest available recombination algorithm: SenoBright\texttrademark ~HD with NIRA giving the best CESM quality recombination images~\cite{gennaro2022artifact}. 

The image annotation was performed by senior radiologists specialized in breast imaging. 
Clinical information on the dataset (age, lesion sizes, time between acquisitions, RECIST 1.1 grade, breast composition, and Background Parenchymal Enhancement (BPE)) are available in Table~\ref{tab:charac}.

\begin{table}[t]
\centering
\begin{tabular}{|rl|c|}
\hline
\multicolumn{2}{|r|}{\textbf{Characteristics}}             & \textbf{Value $\pm$ std} \\ \hline
\multicolumn{2}{|r|}{Age {[}year{]}}                                  &       $43 \pm 8$                         \\ \hline
\multicolumn{2}{|r|}{Average initial lesion size {[}mm{]}}           & $54 \pm 35$               \\ \hline
\multicolumn{2}{|r|}{Average final lesion size {[}mm{]}}             & $23 \pm 30$               \\ \hline
\multicolumn{2}{|r|}{Time between acquisitions {[}days{]}} & $178 \pm 40$                  \\ \hline

RECIST~\cite{eisenhauer2009new}    & Complete response (pCR) & 18/52     \\
                        & Partial response           & 28/52                   \\
                        & Stable disease             & 3/52                    \\
                        & Progressive disease        & 3/52                   \\ \hline

Density      & A - Fatty                  & 5/51                           \\
                        & B - Scattered area         & 27/51                           \\
                        & C - Heterogeneously dense  & 16/51                           \\
                        & D - Extremely dense        & 3/51                            \\ \hline
                        
BPE                  & Minimal           & 9/51                   \\
                                       & Mild              & 34/51                   \\
                                       & Moderate          & 6/51                    \\
                                       & Marked            & 2/51                     \\ \hline
\end{tabular}
\caption{Clinical characteristics of the NAC-CESM cohort (51 patients with 52 lesions)}
\label{tab:charac}
\end{table}

\begin{figure}[t!]
	\centering
        \includegraphics[width=1\textwidth]{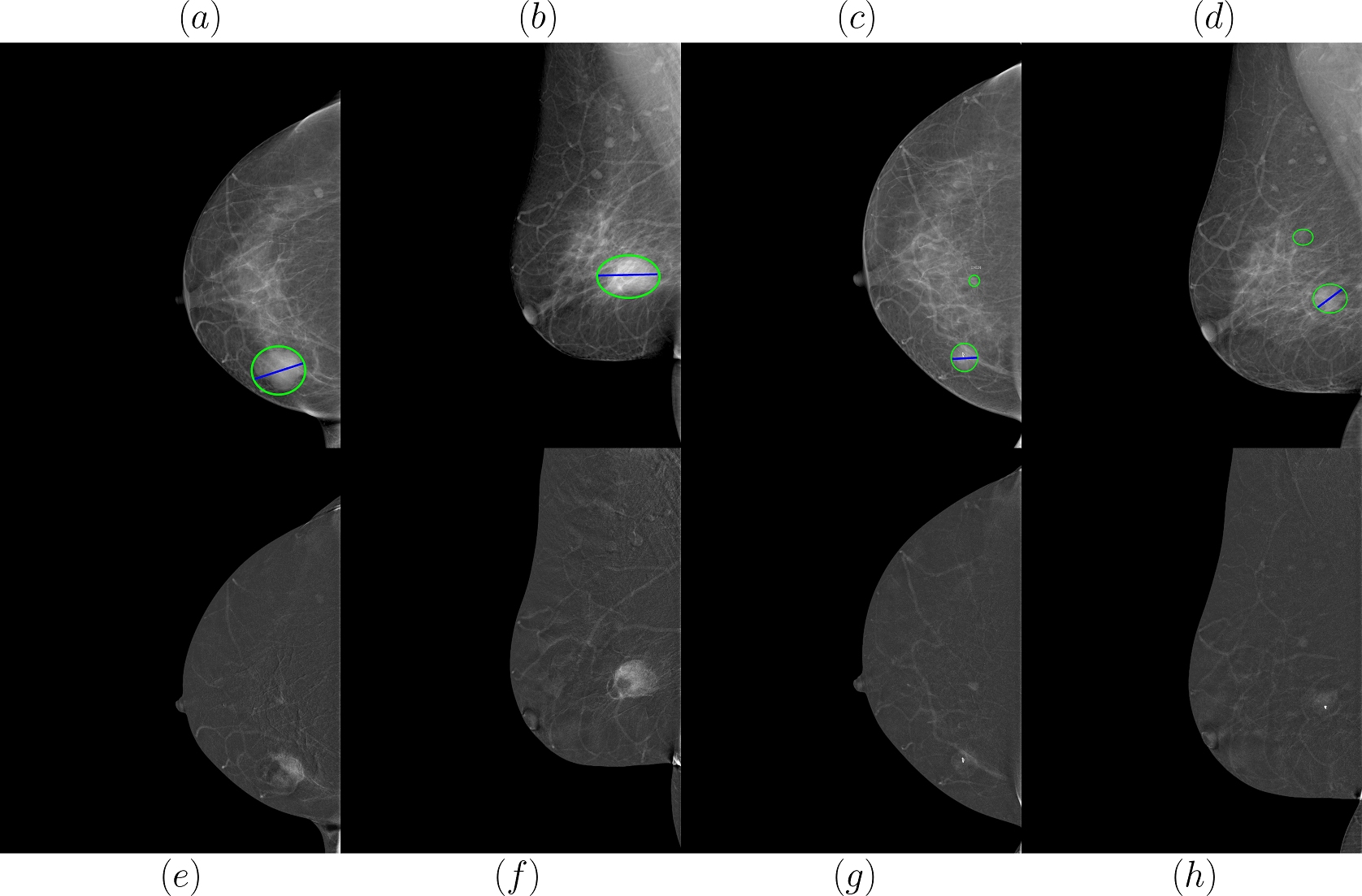}
	\caption{Right breast CESM NAC case with a RECIST: 60\% decrease, partial response. The first line corresponds to low-energy images and the second to the recombined images using NIRA processing. (a) and (b) are respectively the MLO and CC views for the pre-NAC exam, (c) and (d) are the MLO and CC views for the post-NAC exam. The measurements correspond to the lesion extent (green) and length of the largest contrast uptake (blue) used to compute the RECIST. Note that, while displayed on the LE, this measurement is performed both with the LE and recombined image.}
	\label{fig:initial_images} 
\end{figure}

\revision{The acquisition of a patient breast who followed a NAC path} is presented in Figure~\ref{fig:initial_images}. The first line represents the LE acquisitions and the second line the recombined images. The two first columns are pre-NAC acquisitions on the right breast and the two last columns are the post-NAC (respectively CC and MLO views). The lesion, annotated in the LE appears as a contrast uptake in the recombined images. Its size evolves with the treatment and is evaluated \revision{with the RECIST as a partial response. }

% ------------------------------------------------------------------------------
%
%   PART 1 : DATA
%
% ------------------------------------------------------------------------------

% =========================================================
\section{Methods}
% =========================================================

%Registration approaches have been developed in the breast imaging literature.
%Feature-based methods perform registration from extracted local features on the two images (\eg SIFT method). It is difficult to extract consistent features from two mammograms acquired at different dates because of important texture changes.
%Deep learning-based registration techniques~\cite{zhang2019mammographic,haskins2020deep} have been developed in medical imaging. However, the training requires both ground truth and many annotated data that are not available in NAC-CESM.
%Intensity-based registration techniques~\cite{zach2007duality,wedel2009improved} depend on a global pixel intensity matching of the two images. Those methods were applied in mammography registration~\cite{richard2003new,diez2011revisiting,garcia2017similarity} with selected motion model (\eg affine transforms~\cite{rueckert1999nonrigid}, grid with B-splines interpolation~\cite{tortajada2010improving,celaya2015bilateral}).
%The main limits of such approaches appear when the texture intensity highly changes between two images (\eg due to the chemotherapy effect, acquisition parameters changes). 
%Developing a registration method robust to important intensity changes is of utmost importance when dealing with NAC data. 

% ==========================================
% Registration
% ==========================================

\subsection{Registration}

CESM data is composed of a so-called low-energy image, (similar to an FFDM) and a recombined iodine image. The recombined image shows projected iodine content in vascularized tissues. 
With the chemotherapy treatment, a significant evolution of malignant tissues is expected. Registration cannot, therefore, be based on the iodine information. 
In LE images, normal fibro-glandular textures are (globally) conserved with the chemotherapy treatment allowing to follow the internal motion between two exams. 
%Moreover, LE images are very slightly impacted by the iodine content hence an important part of the texture can be registered.

It is here proposed to register LE images (with local intensity corrections) and apply the obtained transformation to the recombined images. 
The registration will deform the pre-NAC image onto the post-NAC image in order to map the initial lesion and breast tumor bed onto the final acquisition.

Given a pair of image, an initial pre-NAC image $f(\bm x)$  and a target post-NAC acquisition $g(\bm x)$, the intensity-based methods define a spatial residual $\rho_u(\bm x)$ that can be written, with $\bm u(\bm x)$ the Lagrangian displacement field. 
\begin{equation}
\rho_u(\bm x, f, g,\bm u)  =
g(\bm x)-f(\bm x+\bm u(\bm x)).
\end{equation}
Note that all registered images were normalized between [0-1].
The sought displacement correction field is the one that minimizes, on the data fidelity term over a spatial region of interest (ROI), a norm (written $\|.\|_\mathcal{N}$) of the residuals $\rho_u$
\begin{equation}
\underset{\bm u}{\Argmin} \{ \| \rho_u(\bm x, f, g,\bm u) \|_\mathcal{N} \}.
\end{equation}
Written as is, the problem is composed of too many degree of freedom and thus is severely ill-posed. Better conditioning is required. Different approaches have been proposed in the medical and material science literature: for example TV-L1 and Global Digital Image Correlation. 

% ==========================================
% TV-L1 registration
% ==========================================
\revision{
\subsubsection{Total Variation}
Total Variation (TV)-L1 is a popular and efficient algorithm for optical flow estimation presented in~\cite{zach2007duality,wedel2009improved,chambolle2011first}}. This approach consists of an L1 data penalty term and total variation regularization, written with $\bm \nabla$ the bi-dimensional gradient operator, and $\lambda$ a penalization coefficient.
\begin{equation}
\underset{\bm u}{\Argmin} \{ \sum_{\bm x \in \text{ROI}} \lambda \cdot | \rho_u(\bm x, f, g,\bm u) | + | \bm\nabla \bm u (\bm x) |  \}.
\end{equation}
This state-of-the-art registration procedure is not adapted to all NAC-CESM registration requirements. This formulation is sensitive to large local intensity changes and thus will not be able to correctly register the lesion area. 

% ==========================================
% Global Digital Image Correlation - GDIC
% ==========================================

\subsubsection{Global Digital Image Correlation - GDIC}
The optical flow conservation approach is an intensity-based registration method also called Digital Image Correlation (DIC)~\cite{sutton2009image}. This approach is widely used in material science~\cite{Grediac} and medical-biomechanical imaging community~\cite{lee2019validation}.

In global DIC~\cite{Besnard06}, written as GDIC in this paper, the motion support is discretized into a space of lower dimension. One solution is to express the motion on a finite element mesh basis. This approach highly reduces the number of degrees of freedom and provides continuous fields of adjustable complexity. Moreover, the finite element mesh can be patient-specific by being adapted to the patient geometry. 

The displacement is written as a 2D mesh kinematics composed of $N_u$ nodes and shape functions $\bm \phi(\bm x)$ for the interpolation. The motion can thus be written 
\begin{equation}
\bm u(\bm x)=\sum_{l=1}^{N_u} u_l\bm \phi_l(\bm x), 
\end{equation}
with $u_l$ the nodal displacements. 
Finally, a classical Newton-Raphson routine is used to minimize the L2 norm
\begin{equation}
\underset{\bm u}{\Argmin} \{  \sum_{\bm x \in \text{ROI}} \| g(\bm x)-f(\bm x+\bm u(\bm x)) \|^2 \}.
\end{equation}
The linearized residual reads 
\begin{equation}
g(\bm x)-f(\bm x+\bm u(\bm x) +\partial  \bm u(\bm x)) \approx g(\bm x)- \tilde f(\bm x)-\partial  \bm u(\bm x) \bm \nabla \tilde f(\bm x),
\end{equation}
with $\tilde f(\bm x)=f(\bm x+\bm u(\bm x))$ the updated warped pre-NAC image. The minimization leads to the following linear system:
\begin{equation}
    [\bm M]\{\partial \bm u\} = \{\bm b\},
    \label{lin1}
\end{equation}
with the correction vector amplitude $\{\partial \bm u\}$, $[\bm M]$ the Hessian of the functional, and $\bm b$ the second member 
\begin{equation}
    M_{ij}=\langle
    S_i,S_j
    \rangle,
\end{equation}
\begin{equation}
    b_{i}=\langle
    S_i,\rho_u
    \rangle.
\end{equation}
$S_i= \bm \phi_{i}\cdot \bm \nabla \tilde f(x)$ is the nodal kinematic sensitivity fields. The notation $\langle\cdot,\cdot\rangle$ denotes the inner product (\ie contraction over  $\bm x \in$ ROI). 

%Note that it is generally expected that, at convergence, the deformed pre-NAC image $\tilde f(\bm x)$ matches the target one $g(x)$. Then, the sensitivity fields can be approximated with its solution at convergence $S_i= \bm \phi_{i}\cdot \bm  \nabla g(x)$ and does not require to be updated at each iteration. Nevertheless, due to the large motion between $f$ and $g$ and due to potential texture evolution, we chose to update $\tilde f(\bm x)$ to simplify the convergence (as the initial state would be far from the converged solution).

Because registered breast images are acquired at different time intervals, the brightness conservation assumption is not respected due to factors extrinsic to the displacement. Those changes may come from evolution in the breast tissue (especially when a NAC treatment is applied), changes in the acquisition system, and parameters (compression force and thickness, tube voltage, exposure time, ...). Those intensity variations cannot be all corrected by a single scalar gain correction (\eg a constant brightness coefficient) but require to be spatially adapted.

% ==========================================
% Relaxed intensity conservation - GDIC-I
% ==========================================

\subsection{Relaxed intensity conservation - GDIC-I}
To consider large local intensity changes in the lesion area an intensity compensation procedure, called GDIC-I, is developed. The approach is based on a brightness contrast correction studied for material science~\cite{mendoza2019correlation}.
The relaxed brightness conservation assumption redefines a correction model for the images, by compensating the intensity variations with a spatial field $v(\bm x)$:
\begin{equation}
\tilde f(\bm x) = f(\bm x+\bm u(\bm x)) + v(\bm x,f(\bm x+\bm u(\bm x))),
\end{equation}
where $v(\bm x, h)$ can be written with a polynomial correction
\begin{equation}
v(\bm x,h(\bm x)) = \sum_p v_p(\bm x) h^p(\bm x).
\end{equation}
In our work, the correction is limited to the second order. The intensity corrected residual, written $\rho_{uv}$ reads
\begin{equation}
\rho_{uv}(\bm x, f, g,\bm u,\bm v)  =
g(\bm x)-\left(
v_0(\bm x) + (1+v_1(\bm x))f(\bm x+\bm u(\bm x))   \right).
\end{equation}
Note that some usual medical X-ray artifacts may require correcting the image with higher orders (\eg beam hardening would require order 3 polynomial~\cite{krumm2008reducing}). 

Similarly to the kinematic model, the intensity correction can also be regularized and written in a finite element framework 
\begin{equation}
v_p(\bm x)=\sum_{l=1}^{N_v} v_{pl} \psi_l(\bm x), 
\end{equation}
with $v_l$ the $N_v$ nodal brightness values and $ \psi(\bm x)$ the intensity interpolation functions.  

All shape functions can be grouped into a single vector of shapes and all degrees of freedom in one vector of parameters. The vectors $\bm \Phi=(\phi,\psi,\psi)$ represent for $\phi$ and $\psi$  respectively the kinematics and the intensity shape functions and $\bm a = (u, v_0, v_1)$ the correction amplitudes.
The linearized problem can be written similarly to eq.\ref{lin1}:
\begin{equation}
    [\bm N]\{\partial \bm a\} = \{\bm n\},
\end{equation}
with $[\bm N]$ the hessian of the functional and $\bm n$ the second member 
\begin{equation}
    N_{ij}=\langle
    S_i,S_j
    \rangle,
\end{equation}
\begin{equation}
    n_{i}=\langle
    S_i,\rho_u
    \rangle.
\end{equation}
$S_k= \Phi_{i}\cdot s_{j}$ is the nodal sensitivity fields (with the super index $k=i\cdot j$ and $j=[1,2,3]$) computed from the image derivatives. 

Finally, at convergence, it can be noted that the residual of clinical interest may still be computed as $\rho_{uv}(\bm x, f, g,\bm u,\bm v=\bm 0)$ that highlights the intensity changes. In the case of NAC, this residual will highlight all textures hence lesion changes.
The proposed intensity corrected approach is designed not to be biased by lesion evolution in NAC. 
In order to successfully register large motions, a regularization and a multi-scale approach are used.

% ==========================================
\paragraph{Regularization}
To make the procedure robust and to avoid possible poor conditioning, a penalization on the functional can be added. While a classical soft Tikhonov regularization~\cite{tikhonov1977solutions} could be used (damping modes with low eigenvalues), it is preferred to introduce a penalization based on the comparison between the estimated displacement field and that of the solution to a homogeneous elastic problem \cite{rethore2009extended}. 
It is important to note that this strategy does not require the studied sample to strictly obey linear elasticity (which is not the case for (i) the breast behavior and (ii) its projected behavior). Rather, it can be seen as a filter that locally dampens abrupt displacement gradients in order to guarantee a smooth and differentiable displacement field. %The unrealistic high localized strains are hence penalized. Such filter can be tuned through the use of a regularization length. 

% ==========================================
\paragraph{Multi-scale approach}
Developing a robust method capable of dealing with all patients in the database is a complex task because of the large range of potential repositioning, texture changes, breast size, \etc
%The breast positioning between two exams may be shifted up to 6~cm~$ ($~600~pixels) with important intensity changes due to the chemotherapy treatment, normal texture change between two exams, and the change in acquisition parameters (breast compressed thickness, beam characteristics, ...). 
A multi-scale approach is thus developed to consider large image differences. This procedure is composed of two multi-scale approaches: an image multi-scale and a model multi-scale.

Image multi-scale: a pyramidal image method that consists in filtering with a Gaussian kernel and down-scaling the images. The registration procedure starts with the filtered scales (composed of the low image frequencies) and progressively retrieves the image details at a finer resolution level. 

Kinematics multi-scale: a model whose complexity evolves with the iterations. A linear projector matrix $\bm A$
is introduced with a kinematic part $\bm A^u$ and an identity matrix of size $N_v$, $\mathbb{I}_{N_v}$:
\begin{equation}
    [\bm A] = \begin{bmatrix}
\bm A^u & \bm 0 \\
\bm 0 & \mathbb{I}_{N_v},
\end{bmatrix}  
\end{equation}
with $N_{\text{red}}$ the number of reduced parameter. $\bm A$ is
of size [$N_u$+$N_v$, $N_{\text{red}}+N_v$] and allows linking the nodal degrees of freedom together thus reducing the effective number of identified kinematics parameters to $N_{\text{red}}+N_v$. The reduced motion field $\bm u_{\text{red}}$ is written:
\begin{equation}
\bm u_{\text{red}}(\bm x)=\sum_{l} \bar u_l \bar{\bm \phi_l} (\bm x) = 
\sum_{l} 
\sum_{i}  u_i A^u_{il}
\sum_{j} \bm \phi_j(\bm x) A^u_{jl}.
\end{equation}
This projector is applied for the computation of a reduced Hessian $\bar N_{kl}$ and second member computation $\bar n_{l}$
\begin{equation}
    \bar N_{kl}=\langle
    S_iA_{ik},S_jA_{jl}
    \rangle,
\end{equation}
\begin{equation}
    \bar n_{l}=\langle
    S_iA_{il},\rho_u
    \rangle.
\end{equation}
As an example, linking all $u_x$ into a single degree of freedom is equivalent to a translation in the $x$ direction. 
Starting with coarse kinematics allows for correcting most of the motion. Then the projector is progressively relaxed until it reaches the identity (\ie all nodes are independent).

The kinematics model starts with rigid body motions (RBM) to coarsely position the breasts, then affine motions are introduced, and finally, the full mesh kinematics can operate with $\bm A^u = \mathbb{I}_{N_u}$, the identity.

% ==========================================
\paragraph{Transformation of CESM images}
After the registration of LE images, recombined pre-NAC image, written $f_\text{Rec}$, can be transformed using the measured displacement field: $f_\text{Rec}(\bm x + \bm u(\bm x))$. The recombined residual field $\rho_{Rec}(\bm x, f_\text{Rec}, g_\text{Rec},\bm u)$ is hence defined with the post-NAC recombined image $g_\text{Rec}$ such as $\rho_{Rec}(\bm x, f_\text{Rec}, g_\text{Rec},\bm u)=g_\text{Rec}(\bm x )- f_\text{Rec}(\bm x+ \bm u(\bm x))$

% ==========================================
% Chosen parameters
% ==========================================

\subsection{Chosen parameters}
The kinematic and intensity meshes were generated from the control of the nodal surface density. This criterion is better than imposing a fixed number of nodes as the breast size varies. The kinematic mesh density was set to $1$~node~/~$5$~cm$^2$ resulting in a median number of $50$ kinematics nodes (standard deviation of 16~nodes) and the intensity mesh distance was set to $1$~node~/~$2$~cm$^2$ resulting in a median number of $120$ nodes (standard deviation of 45~nodes). With this size, the mean lesion size treated with chemotherapy is mapped with approximately 15 and 3 nodes before and after chemotherapy. 
\revision{Discussion on the mesh size and its influence on the registration results is proposed in \ref{annexB}. }

To ensure convergence in all cases, the number of iterations was fixed to 50 with a fixed multi-scales schedule: the first ten iterations are obtained with rigid body motions, the five next with affine transforms and the rest with all free nodes. 

%\begin{table}[h]
%\centering
%\begin{tabular}{|c|cc|c|}
%\hline
%\textbf{Iteration} & \multicolumn{2}{c|}{\textbf{Image scale}}     & \textbf{Model scale} \\ \hline
%1-10                & \multicolumn{1}{c|}{Gaussian Blur 6} & Downsizing 6 & Rigid Body motions         \\ \hline
%10-15  & \multicolumn{1}{c|}{Gaussian Blur 6} & Downsizing 6 & Affine   \\ \hline
%16-30 & \multicolumn{1}{c|}{Gaussian Blur 6} & Downsizing 6 & Identity \\ \hline
%31-40 & \multicolumn{1}{c|}{Gaussian Blur 3} & Downsizing 3 & Identity \\ \hline
%41-50 & \multicolumn{2}{c|}{-}                              & Identity \\ \hline
%\end{tabular}
%\caption{Fixed multi-scale schedule}
%\label{tab:1}
%\end{table}

The TV-L1 approach was applied first with a set of generic hyper-parameters as defined in~\cite{zach2007duality}, written: TV-L1-a, then optimized with a better set of parameters, written TV-L1-b such that (TV-L1-a): $\lambda=15$ and (TV-L1-b): $\lambda=20$.

% ==========================================
% Evaluation procedure
% ==========================================

\subsection{Evaluation procedure}
\label{sec:metrics}

For all the evaluated approaches, a very simple convergence validation was chosen to identify the cases that diverged. Three criteria at the last iteration have to be reached to validate convergence. (1) The residual norm is under the initial residual norm, (2) the residual norm does not vary more than 5\% of the reference image intensity dynamic, and (3) the motion field norm does not vary more than 5\% compared to the previous iteration. When all are respected, the case is labeled as converged.
a
\paragraph{Residual norm:}
First, as it is the minimized metric, the L2 norm of the residuals is used to evaluate the global metric convergence $\sum_{\bm x \in \text{ROI}} \| \rho_{uv}\bm x, f, g,\bm u, \bm v\| ^2$ and assess the global image matching quality. An important residual norm shows an unconverged registration.

\paragraph{Residual field:}
Second, the residual field $\rho_u$ highlights local registration errors (\eg unregistered edges, incorrect registration of the nipple, of the pectoral muscle). 
The visual evaluation of the residual field allows for validating the correction model. 

\paragraph{Landmarks:}
Third, this algorithm aims to be used in a clinical environment. With the goal to locate texture evolution between exams, an evaluation metric is the registration of clinical landmarks. 
618 landmarks, written respectively $\bm x_f$ and $\bm x_g$ have been placed in respectively $f$ and $g$ on clearly identified textures (\eg calcification, nipple, infra-mammary fold, the center of nodules) in the LE and/or the recombined image. The registered landmark distance $d$: $d = \| \bm x_g-\bm x_f-\bm u(\bm x_f)\|_{L_2}$
will be evaluated with simple statistics: Root Mean Square Error (RMSE$[d]$) and maximal error (max$[d]$).

\paragraph{Evaluation of the synthetic case:} 
Because a ground truth is known (both for the lesion size and displacement magnitude) in the synthetic case, two additional metrics can be used. The first is the displacement field comparison: root mean square displacement error, written RMSE$[u]$ (and computed for the $x$ and $y$ direction) and standard deviation, written std$(u)$. The second metric is the quantification of the lesion intensity evolution. This value is obtained when summing the residual intensity on the lesion area.  

\revision{
\paragraph{Presented images/fields:}  
A specific color coding is chosen for each image type and fields allowing (1) to easily identify what image/field the reader is seeing and (2) highlighting specific behavior (e.g. , easy identification of positive and negative patterns).
\begin{itemize}
\item Mammography acquisitions (LE and Rec images). A standard gray level color-map was selected.
\item Residual fields. A divergent color-map (blue/white/red) was selected as it is important to highlight positive and negative patterns.
\item Kinematics field. A hot color-map was selected.
\item Intensity fields. A divergent color-map (brown/white/cyan) was selected as the intensity correction is to be compared to 0.
\end{itemize}
}

% ------------------------------------------------------------------------------
%
%   PART 2 : RESULTS
%
% ------------------------------------------------------------------------------

\section{Results}

% ==========================================
% Results on the synthetic NAC case
% ==========================================

\subsection{Results on the synthetic NAC case}
\begin{figure}[t!]
	\centering
        \subfigure[Displacement field obtained with GDIC]{\includegraphics[width=0.45\textwidth]{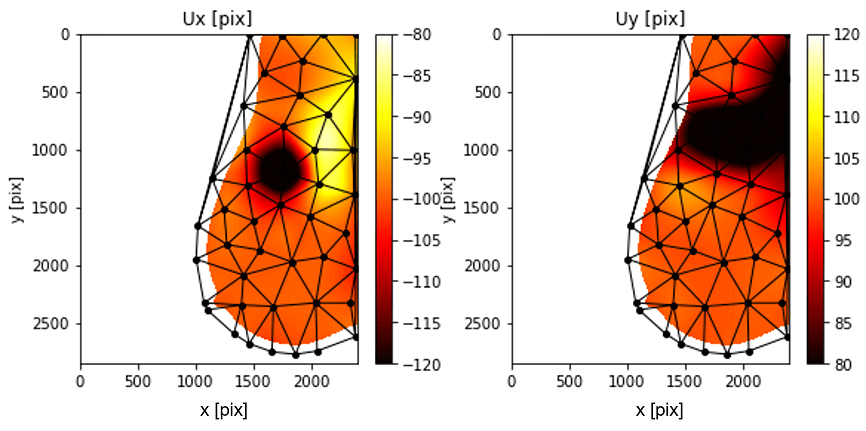}}\\
        \subfigure[Displacement and intensity fields obtained with GDIC-I]{\includegraphics[width=0.45\textwidth]{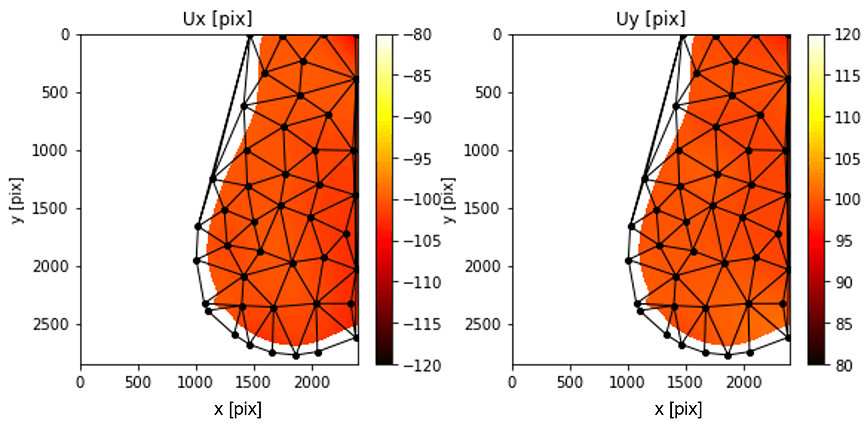}\includegraphics[width=0.48\textwidth]{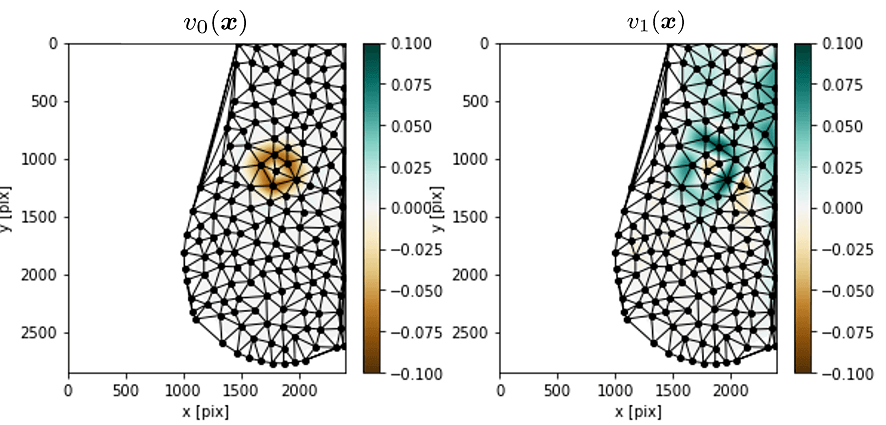}}
	\caption{Outputs of the registration procedure. (i) horizontal displacement $u_x(\bm x)$, (ii) horizontal displacement $u_y(\bm x)$, expressed in pixels (1~pix$\leftrightarrow $0.1~mm) On the second line, GDIC-I fields are also displayed with (iii) $v_0(\bm x)$ and (iv) $v_1(\bm x)$. The kinematics and intensity meshes are shown in black.}
	\label{fig:synthetic2} 
\end{figure}

\begin{figure}[t!]
	\centering
        \subfigure[]{\includegraphics[width=0.24\textwidth]{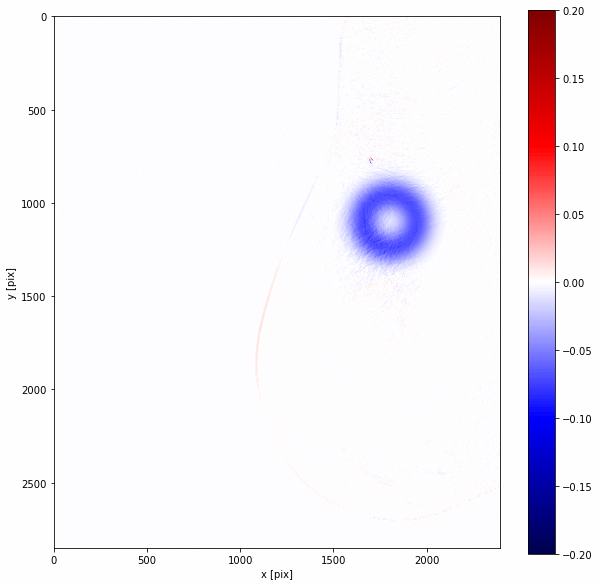}}
        \subfigure[]{\includegraphics[width=0.24\textwidth]{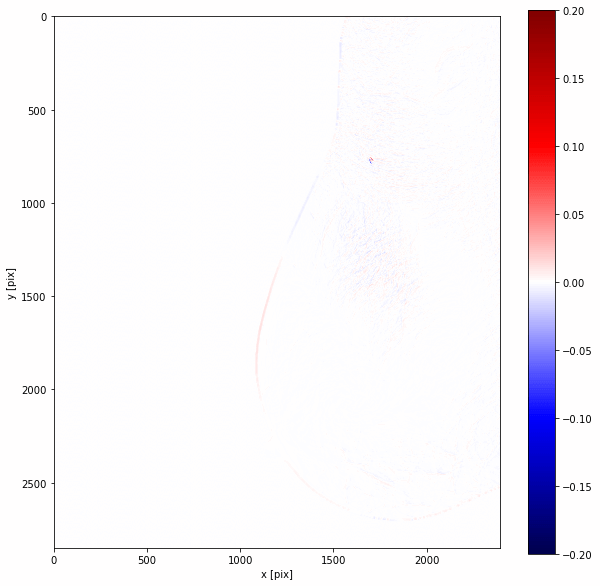}}
        \subfigure[]{\includegraphics[width=0.24\textwidth]{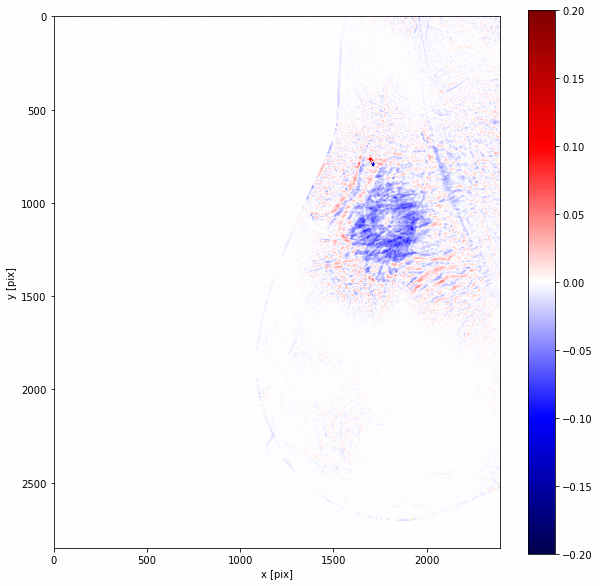}}
        \subfigure[]{\includegraphics[width=0.24\textwidth]{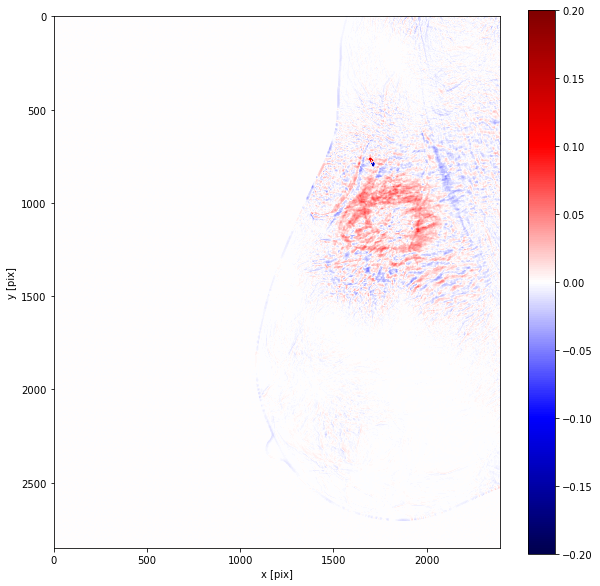}}
	\caption{Residual fields (a) GDIC-I $\rho_{uv}(f,g,\bm u,\bm v=\bm 0)$ (c) GDIC $\rho_{u}(f,g,\bm u)$. The residuals on (b) and (d) have been cleaned from the ground truth lesion evolution $\rho_{uv}(f,g,\bm u,\bm v=\bm 0)-\mathcal{L}_f(\bm x)+\mathcal{L}_g(\bm x)$ and $\rho_{u}(f,g,\bm u)-\mathcal{L}_f(\bm x)+\mathcal{L}_g(\bm x)$.}
	\label{fig:synthetic3} 
\end{figure}

The procedure has been applied to the synthetic case
%The initial residual field is shown in Figure~\ref{fig:synthetic_ini}. 
%
%\begin{figure}[t!]
%	\centering
%        \includegraphics[width=0.3\textwidth]{phantom_ini_res.png}
%	\caption{Initial residual field for the synthetic case.}
%	\label{fig:synthetic_ini} 
%\end{figure}
and the proposed GDIC-I method is compared with GDIC.
At the end of the procedure, the obtained  \revision{displacement and intensity fields are shown Figure~\ref{fig:synthetic2}}. The GDIC-I displacements are much closer to the effective 100 pixels displacement applied to the images than the GDIC fields.
The GDIC-I field $v_0$ shows a circular pattern with an amplitude of 0.1 in the lesion area. This value corresponds to the simulated lesion intensity.
Measured motion metrics are shown in Table~\ref{tab:synthetic}. The RMSE of GDIC-I motion is much smaller in both $x$ and $y$ directions (respectively 0.3 and 0.6~pixels) compared to GDIC (12.1 and 14.2 pixels) and TV-L1-b approaches (33.3 and 7.2 pixels).
\begin{table}[t!]
\centering
\begin{tabular}{c|cc|cc|}
\cline{2-5}
 & \multicolumn{2}{c|}{measured motion in $x$ {[}pix{]}} & \multicolumn{2}{c|}{measured motion in $y$ {[}pix{]}} \\ \hline
\multicolumn{1}{|c|}{} & \multicolumn{1}{c|}{RMSE[$u_x$]} & std($u_x$)  & \multicolumn{1}{c|}{RMSE[$u_y$]} & std($u_y$) \\ \hline
\multicolumn{1}{|c|}{GDIC-I}& \multicolumn{1}{c|}{0.3} &   1.6 & \multicolumn{1}{c|}{0.6}  &   1.6        \\ \hline
\multicolumn{1}{|c|}{GDIC} & \multicolumn{1}{c|}{12.1}  & 9.9 & \multicolumn{1}{c|}{14.2}    & 6.5       \\ \hline
\multicolumn{1}{|c|}{TV-L1-b}    & \multicolumn{1}{c|}{33.3}&   22.1 & \multicolumn{1}{c|}{7.2} &   7.3  \\ \hline
\end{tabular}
\caption{Obtained motion metrics with GDIC-I, GDIC, and TV-L1-b for an input displacement of 100 pixels in $x$ and $y$}
\label{tab:synthetic}
\end{table}
When registered,  residual fields can be recomputed without the intensity term $\rho_{uv}(f,g,\bm u,\bm v=\bm 0)$. Those residual fields, that aim at highlighting all texture changes are shown Figure~\ref{fig:synthetic3}. A blue area corresponds to an intensity decrease and a red area to an intensity increase. The obtained annulus shape is expected as it is the true difference between the two lesions (pre and post-NAC).
Because the ground truth is known, it can be subtracted to the residual field to isolate only residuals due to incorrect motion measurement (Figure~\ref{fig:synthetic3} (b) and (c)).

The sum of intensities of the residual field in the area of the lesion can be performed. This integral shows intensity evolution between two acquisitions. The obtained values are compared to the ground truth evolution. The lesion intensity evolution error with GDIC-I, GDIC, and TV-L1-b are respectively 0.7~\%, 34.2~\%, and 11.3\%. 

%With this synthetic case it can be concluded that the intensity change in the lesion area impacts and biases the registration results. The use of the proposed intensity correction method allows for identifying the correct repositioning and breast texture evolution in time. 

%Finally, some simple quantitative metrics showed that the lesion evolution was well represented when the proposed intensity correction was applied. Without the intensity compensation, the motion partially corrects the lesion change. This imperfect correction is difficult to exploit and hides the texture change in the residual field. 
%When considering local intensity corrections, all the information can be read and quantified from the residual image.

% ==========================================
% Results on clinical data
% ==========================================

\subsection{Results on clinical data}

% ==========================================

\paragraph{Results on one specific NAC case}

\begin{figure}[h]
	\centering
        \includegraphics[width=1\textwidth]{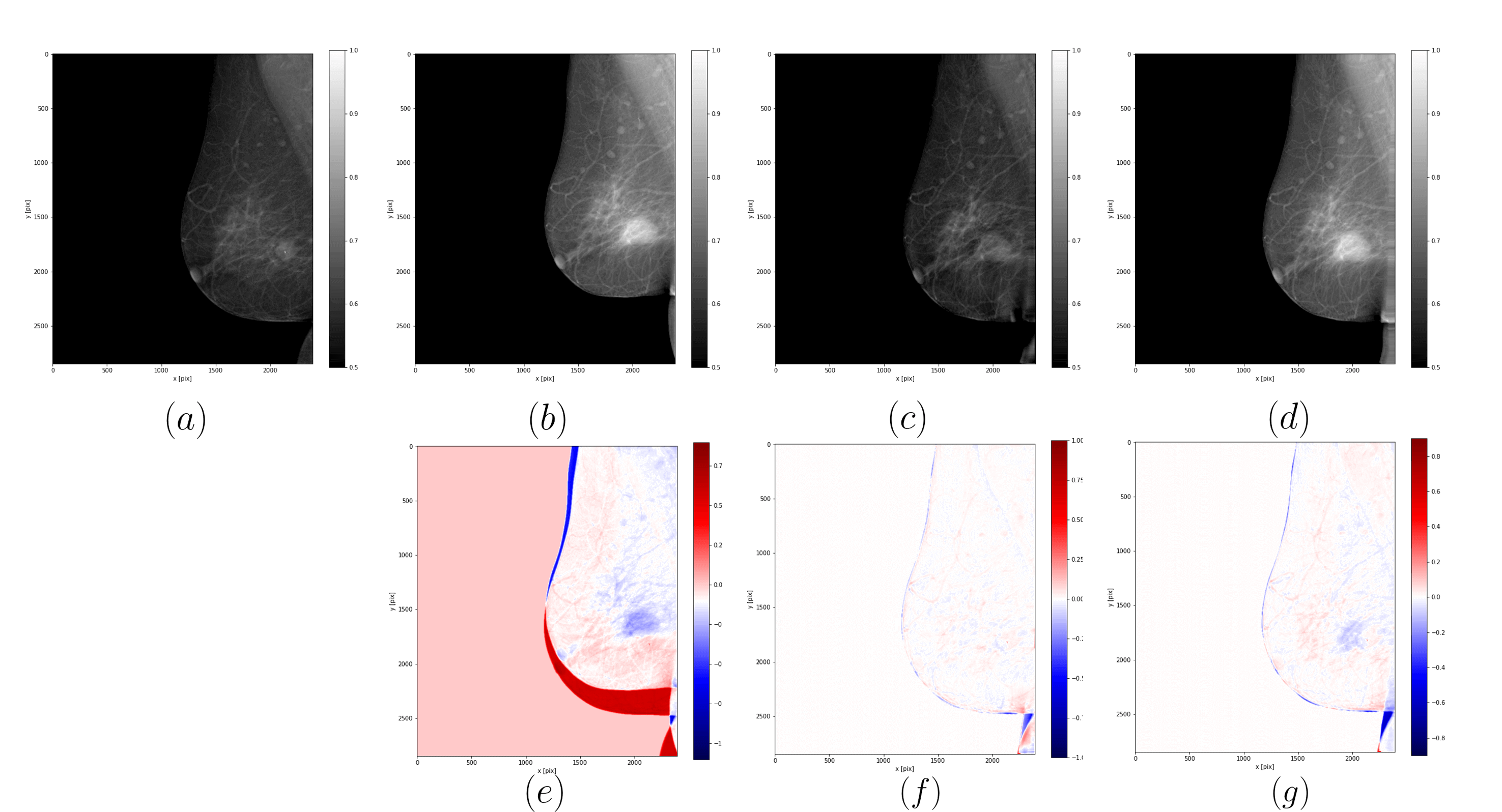}
	\caption{Case of a partial radiological response. First line: low-energy images, (a-b) corresponds respectively to acquired post-NAC RMLO, pre-NAC RMLO, (c-d) are transformed pre-NAC images, respectively with and without applying intensity correction fields. The second line corresponds to the initial residual fields composed of high intensity differences: (e) $\rho_{uv}(f,g,\bm u=\bm 0, \bm v=\bm 0)$, (f) $\rho_{uv}(f,g,\bm u, \bm v)$ and (g) $\rho_{uv}(f,g,\bm u, \bm v=\bm 0)$. } 
	\label{fig:initial} 
\end{figure}

Figure~\ref{fig:initial} presents registration results for one LMLO case showing a complete pathological response (pCR). The GDIC-I registered pre-NAC image with respect to the post-NAC images are shown in Figure~\ref{fig:initial}(c) with intensity correction and (d) without applying the intensity correction field. 
Figure~\ref{fig:initial}(f) represents registered residual fields with intensity correction $\rho_{uv}(f,g,\bm u, \bm v)$. When registered, residual intensities are much smaller than the initial ones. No strong residuals are located on the edges. 
Figure~\ref{fig:initial}(g) shows the the final residual fields without applying the intensity correction field $\rho_{uv}(f,g,\bm u, \bm v= \bm 0)$. 

\begin{figure}[t!]
	\centering
        \includegraphics[width=1\textwidth]{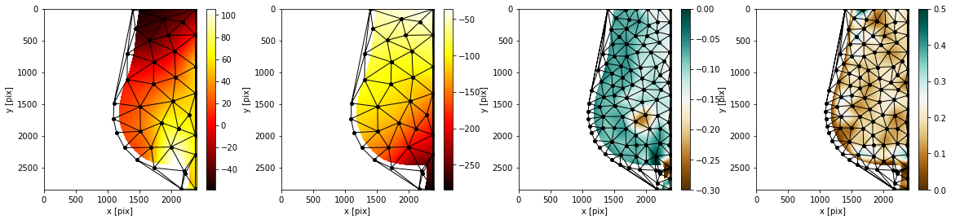}\\
        (a) $u_x(\bm x)$ \hspace{2.5cm} (b) $u_y(\bm x)$ \hspace{2.5cm} (c) $v_0(\bm x)$ \hspace{2.5cm} (d) $v_1(\bm x)$
	\caption{Outputs of the registration procedure for the current/prior registration of one patient. (a-b) horizontal displacement and vertical displacement expressed in pixels (1~pix$\leftrightarrow $0.1~mm) and intensity corrections (c-d). The kinematics and intensity meshes are shown in black.}
	\label{fig:ouputs}  
\end{figure}

% ---------------
The displacement and intensity correction fields are shown in Figure~\ref{fig:ouputs}. An important motion has been captured (more than 280 pixels) with the multi-scale procedure. This motion is quite smooth (maximal absolute strain in $xx$, $yy$ and $xy$ (shear) are respectively 18\%, 10\% and 0.6\%).
The intensity correction fields show an important decrease of intensity in the central lesion area (decrease of approximately 30\%). 

\begin{figure}[t!]
	\centering
        \includegraphics[width=0.75\textwidth]{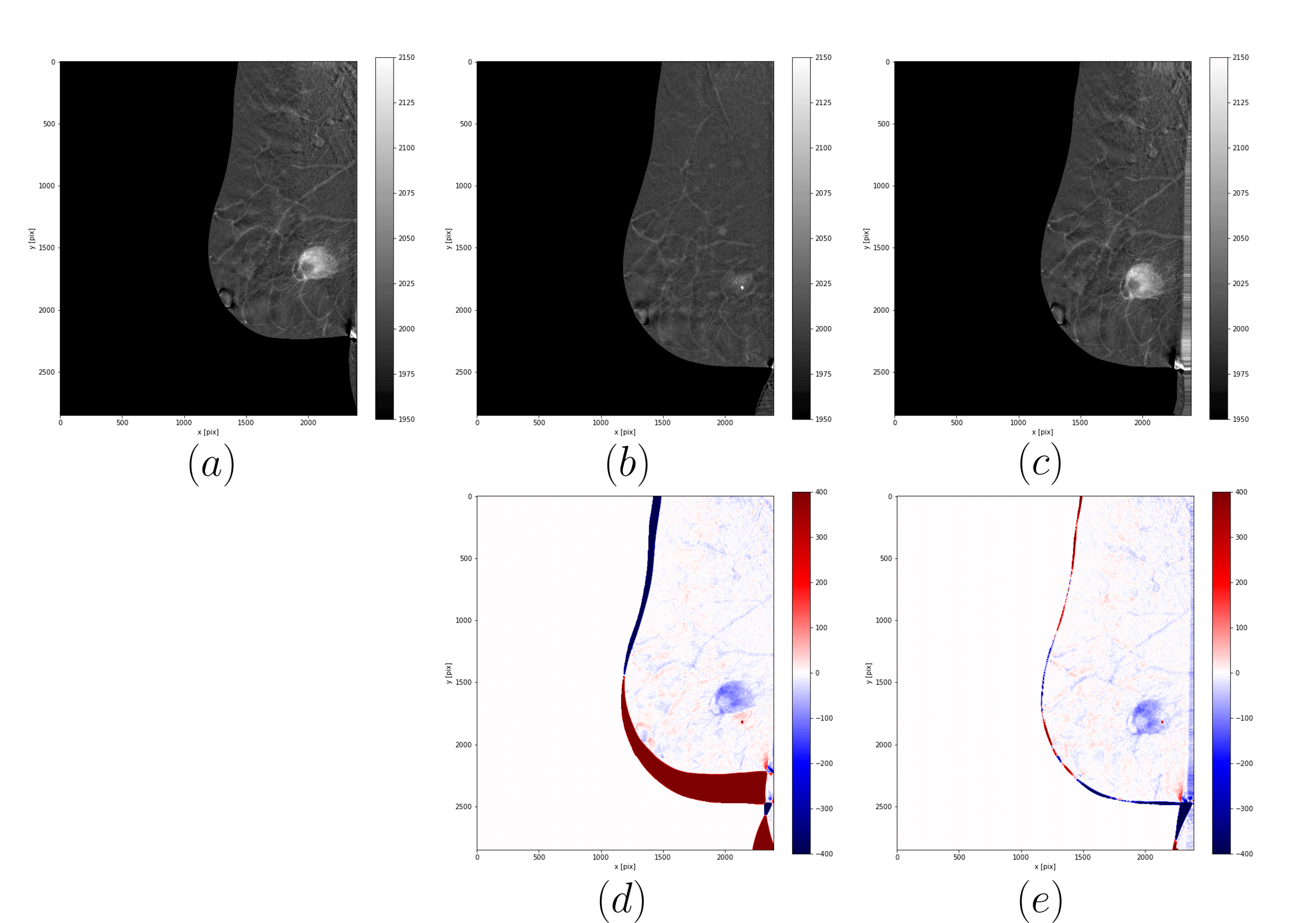}
	\caption{The first line corresponds to recombined images (a) pre-NAC $f_\text{Rec}$ and (b) post-NAC $g_\text{Rec}$ and (c) transformed pre-NAC. The second lines shows recombined residual fields (d) $\rho_{Rec}(f_\text{Rec},g_\text{Rec},\bm u=\bm 0)$ and (e) $\rho_{Rec}(f_\text{Rec},g_\text{Rec},\bm u)$}
	\label{fig:cesm}  
\end{figure}

Pre-NAC, post-NAC and transformed post-NAC are shown Figure~\ref{fig:cesm} respectively in Figure~(a), (b) and (c). The initial and final recombined residuals are shown in (d) and (e). From the assumption made earlier that recombined images display projected iodine quantity, red and blue intensities highlight respectively an increase and decrease of projected iodine between the pre and post-NAC exams.

% ==========================================

\begin{figure}[t]
	\centering
        \includegraphics[width=0.9\textwidth]{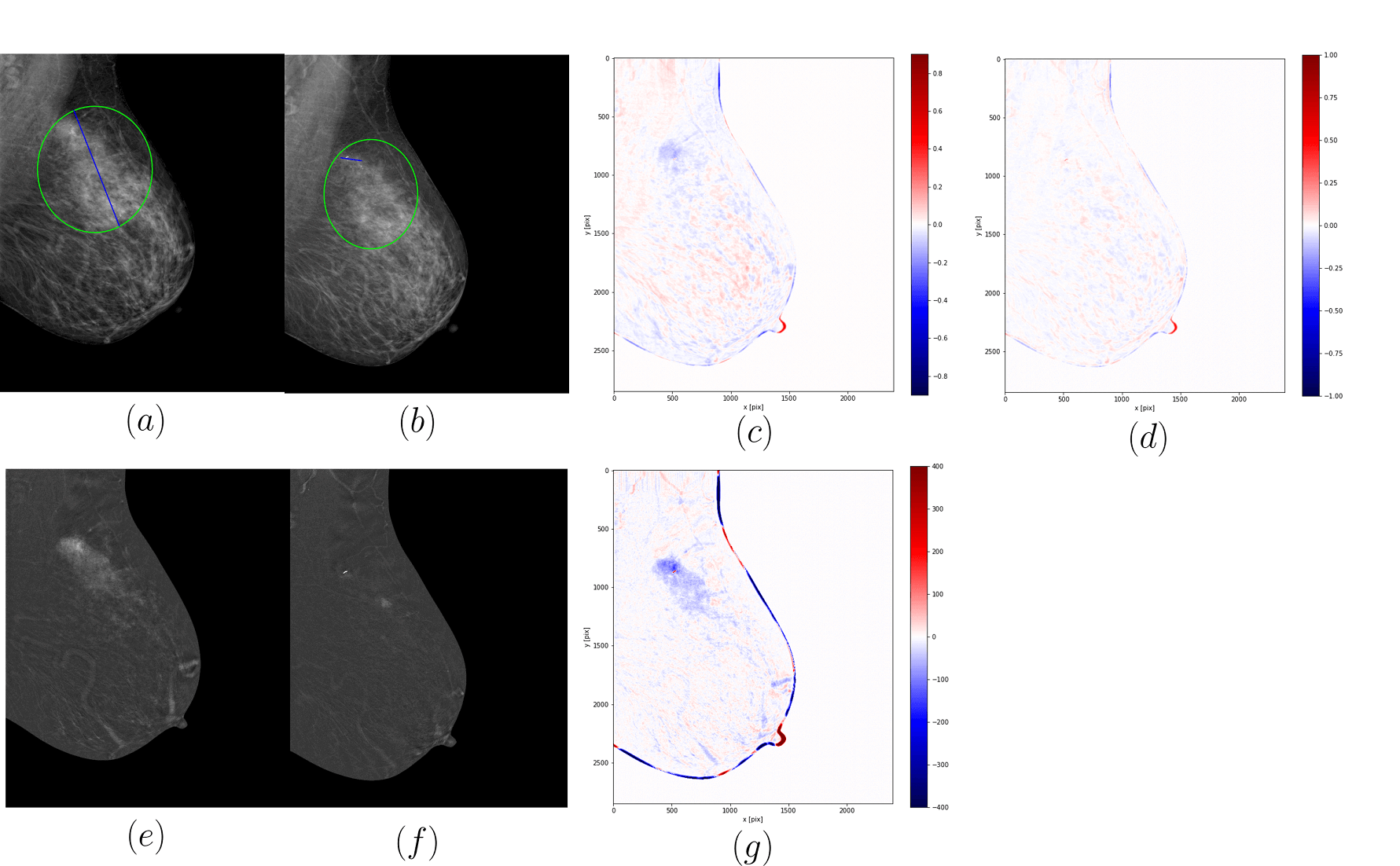}
	\caption{Case of a  partial response (RECIST: 83\% decrease). The first line corresponds to low-energy images with (a) the pre-NAC, (b) the post-NAC, (c) $\rho_{uv}(f,g,\bm u, \bm v=\bm 0)$ and (d) $\rho_{uv}(f,g,\bm u, \bm v)$.
	The second line corresponds to recombined images with (e) the pre-NAC, (f) the post-NAC, (c) $\rho_{Rec}(f_\text{Rec},g_\text{Rec},\bm u)$.}
	\label{fig:res3}  
\end{figure}

\begin{figure}[t!]
	\centering
        \includegraphics[width=0.9\textwidth]{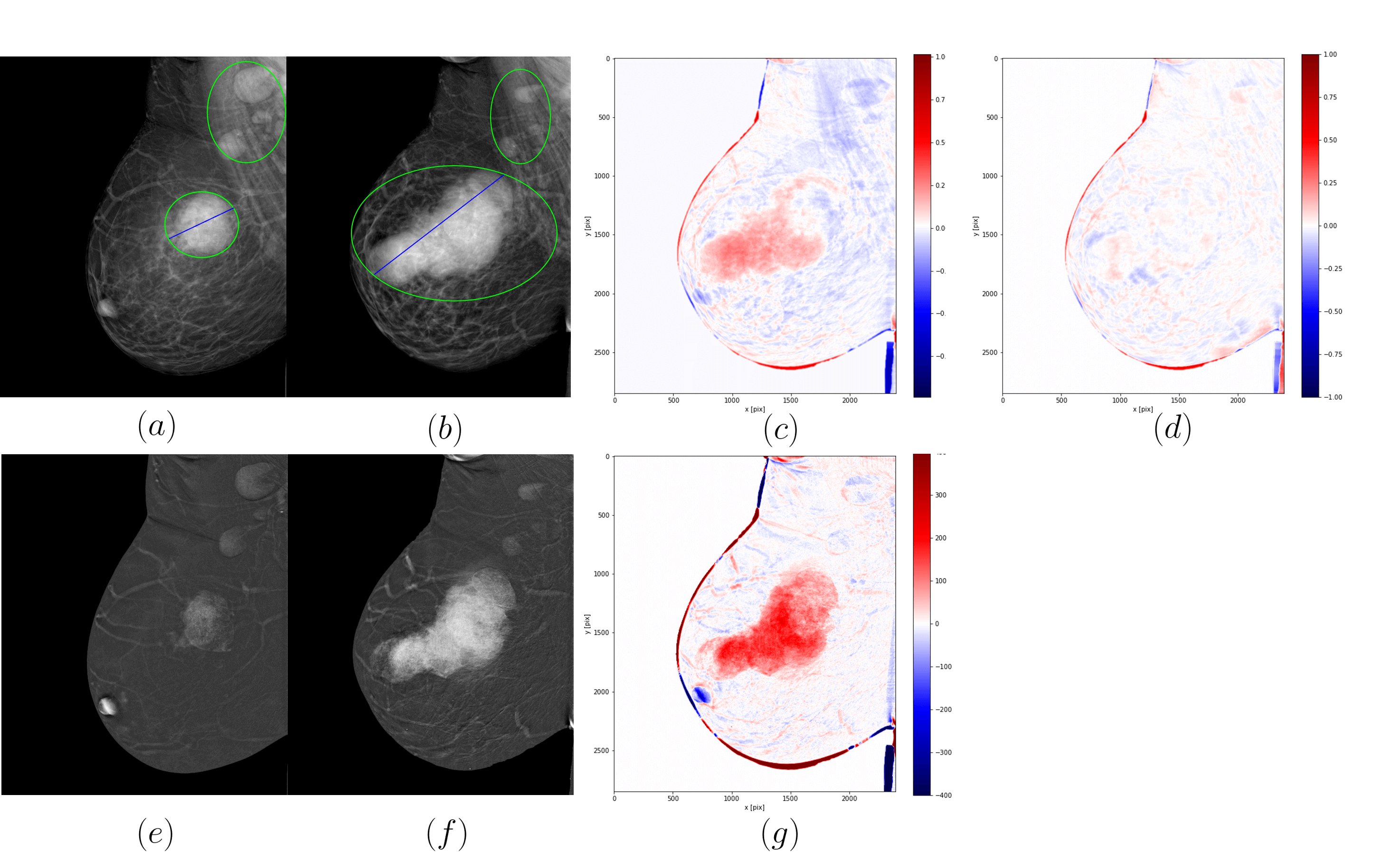}
	\caption{Case of a progressive radiological response (RECIST = 240\%). The first line corresponds to low-energy images with (a) the pre-NAC, (b) the post-NAC, (c) $\rho_{uv}(f,g,\bm u, \bm v=\bm 0)$ and (d) $\rho_{uv}(f,g,\bm u, \bm v)$.
	The second line corresponds to recombined images with (e) the pre-NAC, (f) the post-NAC, (c) $\rho_{Rec}(f_\text{Rec},g_\text{Rec},\bm u,)$.}
	\label{fig:res5}  
\end{figure}

\paragraph{Results on other cases}
Convergence was achieved for every cases (208/208) with the GDIC-I approach. Two cases gave high final residuals because the registration was performed with two highly changed views. One of these challenging cases is discussed in this document.
The numbers of converged cases with different methods are shown in Table~\ref{tab:converged}. A data split shows converged cases when the lesion appears in the image and when no lesion appears (\eg, normal breast, CC view with a lesion in the axillary area). The proposed approach limited to rigid body motion also converged in all cases (208/208). The state-of-the-art techniques, TV-L1-a and TV-L1-b, converged in significantly fewer cases: 121/208 and 169/208 respectively. 
\begin{table}[t]
\centering
\begin{tabular}{l|c|c|c|}
\cline{2-4}
 &
  \begin{tabular}[c]{@{}c@{}}Converged cases with\\  apparent lesion\end{tabular} &
  \begin{tabular}[c]{@{}c@{}}Converged cases without \\ apparent lesion\end{tabular} &
  Total \\ \hline
\multicolumn{1}{|l|}{GDIC-I}            & 101/101 (100\%) & 107/107 (100\%) & 208/208 (100\%) \\ \hline
\multicolumn{1}{|l|}{RBM}               & 101/101 (100\%) & 107/107 (100\%) & 208/208 (100\%) \\ \hline
\multicolumn{1}{|l|}{GDIC}              & 88/101 (87\%) & 95/107 (89\%)   & 183/208 (88\%) \\ \hline
\multicolumn{1}{|l|}{TV-L1-a}           & 39/101  (39\%) & 82/107 (76\%)  & 121/208 (58\%) \\ \hline
\multicolumn{1}{|l|}{TV-L1-b}           & 75/101  (75\%) & 94/107 (89\%)  & 169/208 (81\%) \\ \hline
\end{tabular}
\caption{Converged cases with different registration methods.}
\label{tab:converged}
\end{table}
Results are shown for different RECIST evaluations (Figure~\ref{fig:res3} and Figure~\ref{fig:res5}). 
In those figures, the first line corresponds to  LE images (a-b) and intensity compensated residuals (c-d). Post-NAC recombined images (e-f) are then transformed using the displacement field measured between LE images. The recombined residual field is displayed in (g).
%From the intensity corrected LE residual fields in images (c) and (d), it can be seen in all the cases that the breast edges and inner texture are well registered. The residual field obtained without applying the intensity correction shows the texture intensity evolution. This tissue change may not be clearly identified when looking only at the LE images. The analysis of the residual field is an easier way to analyse the lesion evolution.When computing the recombined residuals, the iodine difference is clearly visible in red/blue colors. %The blue area that appears in the lower quadrant is the major intensity decrease and corresponds to the lesion extent.

The case in Figure~\ref{fig:res5} is a progressive disease and hence has a lesion more intense in the post-NAC than in the pre-NAC. This intensity increase appears in dark red. The difference in LE shows an intensity increase in the direction of the nipple. 

% ==========================================

\begin{figure}[t!]
	\centering
        \includegraphics[width=0.45\textwidth]{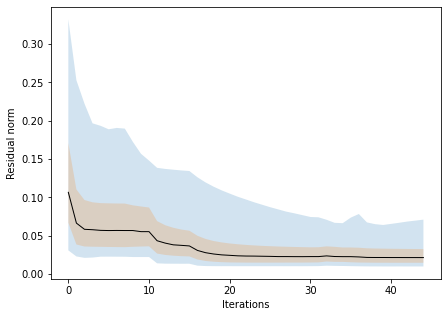}
	\caption{Blue: envelope of the 208 L2 norms of the registered LE residuals for all computations. Orange: mean +/- standard deviation. The black line represents the median values.}
	\label{fig:conv}  
\end{figure}

\paragraph{Evaluation metrics}
The residual norm of all cases at each iteration is presented in Figure~\ref{fig:conv}. 
The median convergence curve is shown with a plain black line. Starting from various initial errors (depending on the repositioning displacement state), the norm decreases with multiple plateaus for each set of optimization parameters. 
The two cases with important breast areas out of the field of view are the ones with the major residual norm.

\begin{figure}[ht!]
	\centering
        \includegraphics[width=1.\textwidth]{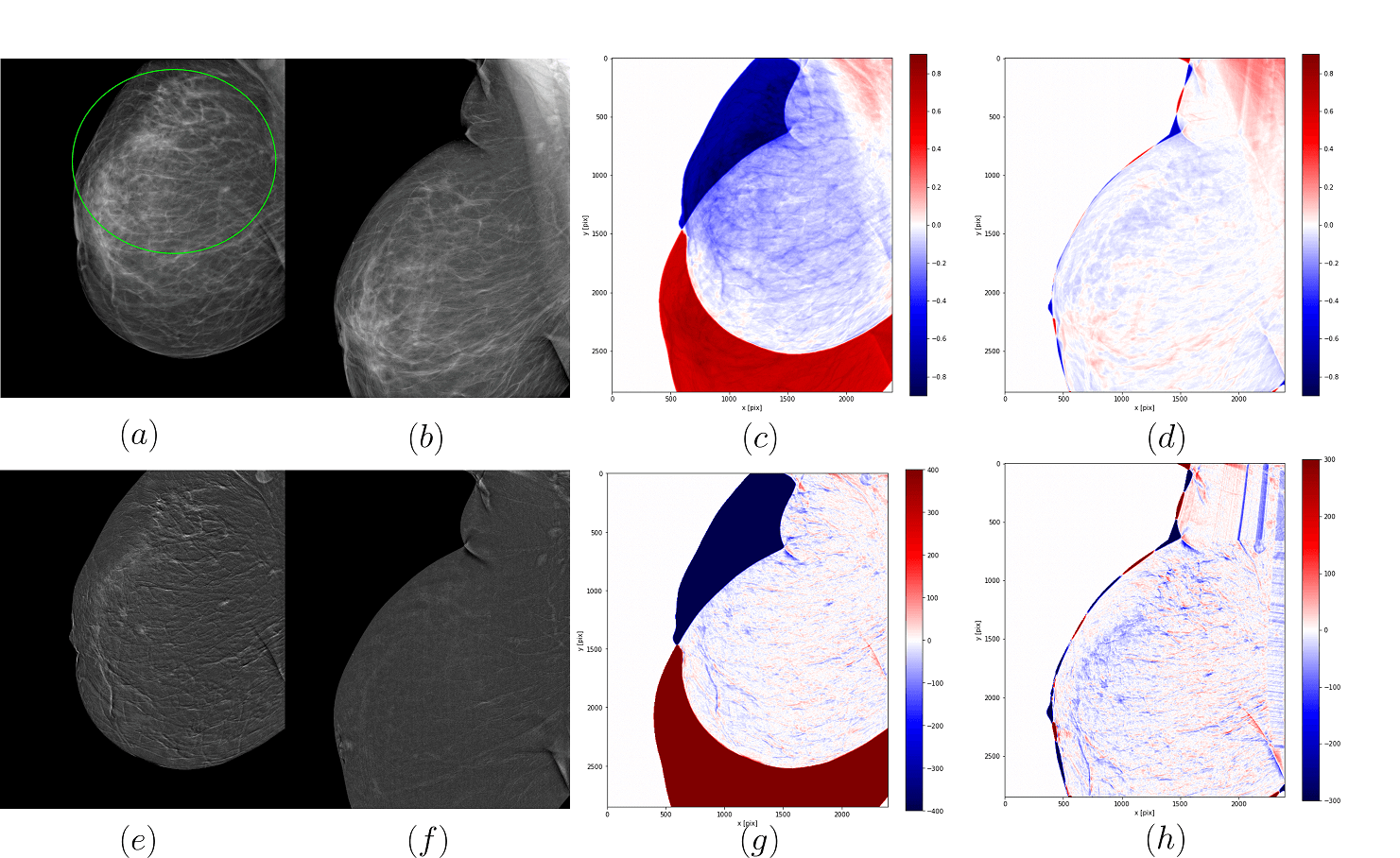}
	\caption{Registered challenging case, using GDIC-I, with an important breast area out of the field of view. (a-b) are the pre and post-LE images without the intensity compensation. (c) and (d) are the initial and registered residuals. The second line shows the Recombined images.}
	\label{fig:challenge}  
\end{figure}

The registration of one of those challenging cases is shown in Figure~\ref{fig:challenge}. An important area of the breast is out of the field of view in the post-NAC image. The registration is challenging with a displacement of more than 6 cm. 

The comparison of the landmarks registration is presented in Table~\ref{tab:tab2}. This table presents the RMSE and maximum distance for  registered landmarks using different approaches. The values were computed only on converged cases. RBM, TV-L1-a, and TV-L1-b have metrics in the same range (RMSE around 10-11~mm). GDIC and GDIC-I led to lower RMSE values, respectively 8.1~mm and 5.9~mm.

\begin{table}[t!]
\centering
\begin{tabular}{c|c|c|c|c|c|c|}
\cline{2-7}
 &
  Initial &
  RBM &
  \begin{tabular}[c]{@{}c@{}}TV-L1-a \end{tabular} &
  \begin{tabular}[c]{@{}c@{}}TV-L1-b \end{tabular} &
  GDIC &
  GDIC-I \\ \hline
\multicolumn{1}{|c|}{RMSE$[d]$ {[}mm{]}} &
  19.0 &
  10.8 &
  11.2 &
  10.8 &
  8.1 &
  5.9 \\ \hline
\multicolumn{1}{|c|}{Max$[d]$ {[}mm{]}} &
  60.6 &
  39.1 &
  55.1 &
  46.2 &
  32.2 &
  21.3 \\ \hline
\end{tabular}
\caption{Registration of clinical landmarks. Note that, for TV-L1 results, values are computed only for the cases that converged.}
\label{tab:tab2}
\end{table}

\section{Discussion}

% ==========================================
% Discussion on the synthetic case
% ==========================================

\subsection{Discussion on the synthetic case}
Because of important motion, the initial residual shows important positive and negative values on the edges, on the lesion area, and the pectoral muscle. It can be noted that the blue shape on the lesion area is very irregular and cannot be quantitatively interpreted.  
 
Figure~\ref{fig:synthetic2}(a) shows that GDIC is highly affected by the lesion evolution. Instead of a constant field, an important compression field aims at reducing its size so it better matches with the post-NAC image.
On the other side, GDIC-I displacement is much smoother and close to the ground truth of 100~pixels. This field is not affected by the lesion change. The obtained intensity fields compensate for the lesion evolution.  The metrics of the displacement for GDIC-I are much better than GDIC. Although only $v_0$ should be activated if the intensity correction support was perfect, it can be noted that the fields $v_0$ and $v_1$ are coupled. 

When converged, GDIC-I residual field highlights the lesion change as it disappears when removing the ground truth lesion change. The GDIC-I residual is then a quantitative description of the texture evolution as it is cleaned from the motion impact.
On the opposite, the GDIC residual field does not have a smooth circular shape. This field does not correspond to the ground truth lesion change and cannot be interpreted quantitatively.

It can be noted that very similar observations on displacement field were obtained from a real NAC case displayed in \ref{annexA}.

The intensity compensation appears to be important when performing the registration. It allows quantitative information (both in residuals and displacement fields) to be read.

% ==========================================
% Discussion on the clinical cases
% ==========================================

\subsection{Discussion on the clinical cases}

The proposed method achieved a high convergence rate as all cases in our database converged. 
The number of iterations is intentionally large to maximize registration quality instead of computation time. Designing a stopping criterion based on the residual value or evolution could be envisioned to speed up the computation.
In the two cases with important motion (more than 6 cm), the registration successfully matched the breast edges. However, the inner texture may not be perfectly registered (\eg important 3D repositioning motion, fibro-glandular texture changes).
On all metrics, the proposed approach outperformed other state-of-the-art approaches. %This validates the robustness requirement for clinical application.
For the TV-L1 and GDIC methods, the ratio of converged cases without an appearing cancer is significantly higher than the cases where the lesion is visible. This comes from the high impact of the cancer intensity changes. From the landmarks measurements, it can be seen that the proposed GDIC-I approach leads to a much smaller registration distance error (RMSE of 5.9~mm).

What remains in the clinical residual fields can be split into different categories: %(discussed in annex A):
\begin{itemize}
    \item A physical texture difference. This difference corresponds to fibroglandular evolution, lesion evolution, iodine changes, etc.
    \item A 3D motion effect that cannot be corrected by 2D in-plane procedure. The proportion of this residual may be important when the breast is very dense. 
    %This type of residual is discussed in Annex A.
    \item Noise and artifacts that cannot be registered.
\end{itemize} 
In some residuals, it can be seen a vertical pattern on the image borders. This is due to a constant intensity extrapolation pulling the breast texture inside the field of view.  Note that this pattern could be masked in clinical practice.

In general, it can be seen that the low-energy residual gives a piece of new information on the intensity evolution. 
It is well correlated with the iodine residual and with the lesion annotation (represented in the LE images with the green ellipse). Lesion intensity evolution that may be difficult to read in the LE images can be more easily identified.
Its reading is fast and enables to efficiently detect and quantify the texture changes, lesion position, and extent. This field could be used for later clinical discussion with the surgeon and/or the oncologist.

It can be noted from the registered residuals that the biopsy clips were inserted in the center of the pre-NAC lesions. The clip appears as a small dark red area as it is not seen in the pre-NAC image. In a quantitative analysis, this clip could be erased to focus on iodine evolution content.

It can be noted that LE residual norm is related to the patient breast texture. A very glandular breast with important texture change generally has a more important residual field than a fatty breast. The proposed approach will have lower accuracy on very dense breasts (ACR-D).

When applying the motion correction to the recombined image, the image difference shows iodine evolution. It is hence important to use the best available recombination algorithm not to be polluted by artifacts. 

The computation time is approximately 1 min per registered pair of images on a computer with a configuration similar to the acquisition system and a script in python. 
With great opportunities of computational improvement (\eg implementing a stopping criterion, writing the script in a more efficient framework, \revision{and using GPU programming~\cite{ostergaard2008acceleration}}), it is an important perspective to speed up the procedure for clinical usage.

% ------------------------------------------------------------------------------
%
%   PART 3:  CONCLUSION
%
% ------------------------------------------------------------------------------

\section{Conclusion}

A robust non-rigid registration method, GDIC-I, able to handle large motion, and important intensity changes with a high convergence rate has been proposed. 
The approach consists in generating smooth intensity fields compensating for local texture changes. After registration is performed, this intensity compensation field is removed highlighting the lesion changes.
First applied on a synthetic example where a ground truth is known, the method was executed on 208 CESM low-energy image pairs. Cleaned from its motion, the image difference allows identifying breast texture changes thus tumor response. 
After motion identification, the transformation field was applied to the recombined images. With the same principle, the recombined image difference displayed iodine evolution.
 
The proposed approach was compared to the standard state-of-the-art registration techniques. Those classical methods are very sensitive to large breast texture changes and fail when registering NAC cases with large lesion evolution.
Different metrics were used and showed better results with the proposed approach.
The residual field is a piece of interesting clinical information that reveals texture changes. Some NAC evaluations require catching very slight lesion extent evolution, especially in LE images. 
\revision{The residual map may give additional clinical help to the radiologist.  
In those difference map, the radiologist may identify the lesion shrinkage pattern with the treatment, the mapping of the pre-NAC lesion in the post-NAC view for surgical procedures, the identification of the lesion center in post-NAC complete responses, \etc}

In addition to its visual interest, a perspective of this work is to include this registration in a quantification tool for lesion evolution. Proven to be efficient in MRI quantification, it can also be investigated for lesion evolution quantification in CESM.

% ------------------------------------------------------------------------------
%
%   PART 3:  Acknowledgements
%
% ------------------------------------------------------------------------------

\section{Acknowledgements}

The authors would like to acknowledge Asmaa Alaa Eldin Shafy Anany (Baheya Center), Razvan Iordache (GE Healthcare), and Jean-Paul Antonini (GE Healthcare) for their help in the data collection. 

\section*{Compliance with Ethical Standards}
\label{sec:ethics}
This research study was conducted retrospectively using anonymized human subject data made available by \emph{Baheya Foundation For Early Detection And Treatment Of Breast Cancer, Giza, Egypt}. Applicable law and standards of ethic have been respected.

\section*{References}
\bibliographystyle{unsrt}
\bibliography{Biblio} 

\begin{thebibliography}{10}

\bibitem{fowler2017imaging}
Amy~M Fowler, David~A Mankoff, and Bonnie~N Joe.
\newblock Imaging neoadjuvant therapy response in breast cancer.
\newblock {\em Radiology}, 285(2):358--375, 2017.

\bibitem{ou2015deformable}
Yangming Ou, Susan~P Weinstein, Emily~F Conant, Sarah Englander, Xiao Da,
  Bilwaj Gaonkar, Meng-Kang Hsieh, Mark Rosen, Angela DeMichele, Christos
  Davatzikos, et~al.
\newblock Deformable registration for quantifying longitudinal tumor changes
  during neoadjuvant chemotherapy.
\newblock {\em Magnetic resonance in medicine}, 73(6):2343--2356, 2015.

\bibitem{liu2019radiomics}
Zhenyu Liu, Zhuolin Li, Jinrong Qu, Renzhi Zhang, Xuezhi Zhou, Longfei Li, Kai
  Sun, Zhenchao Tang, Hui Jiang, Hailiang Li, et~al.
\newblock Radiomics of multiparametric {MRI} for pretreatment prediction of
  pathologic complete response to neoadjuvant chemotherapy in breast cancer: a
  multicenter study.
\newblock {\em Clinical Cancer Research}, 25(12):3538--3547, 2019.

\bibitem{jahani2019prediction}
Nariman Jahani, Eric Cohen, Meng-Kang Hsieh, Susan~P Weinstein, Lauren
  Pantalone, Nola Hylton, David Newitt, Christos Davatzikos, and Despina
  Kontos.
\newblock Prediction of treatment response to neoadjuvant chemotherapy for
  breast cancer via early changes in tumor heterogeneity captured by
  {DCE}-{MRI} registration.
\newblock {\em Scientific reports}, 9(1):1--12, 2019.

\bibitem{fan2021radiomics}
Ming Fan, Hang Chen, Chao You, Li~Liu, Yajia Gu, Weijun Peng, Xin Gao, and
  Lihua Li.
\newblock Radiomics of tumor heterogeneity during longitudinal dynamic
  contrast-enhanced magnetic resonance imaging for predicting response to
  neoadjuvant chemotherapy in breast cancer.
\newblock {\em Frontiers in molecular biosciences}, 8:119, 2021.

\bibitem{riyahi2018quantifying}
Sadegh Riyahi, Wookjin Choi, Chia-Ju Liu, Hualiang Zhong, Abraham~J Wu, James~G
  Mechalakos, and Wei Lu.
\newblock Quantifying local tumor morphological changes with jacobian map for
  prediction of pathologic tumor response to chemo-radiotherapy in locally
  advanced esophageal cancer.
\newblock {\em Physics in Medicine \& Biology}, 63(14):145020, 2018.

\bibitem{riyahi2018quantification}
Sadegh Riyahi, Wookjin Choi, Chia-Ju Liu, Saad Nadeem, Shan Tan, Hualiang
  Zhong, Wengen Chen, Abraham~J Wu, James~G Mechalakos, Joseph~O Deasy, et~al.
\newblock Quantification of local metabolic tumor volume changes by registering
  blended pet-ct images for prediction of pathologic tumor response.
\newblock In {\em Data Driven Treatment Response Assessment and Preterm,
  Perinatal, and Paediatric Image Analysis}, pages 31--41. Springer, 2018.

\bibitem{wodzinski2018improving}
Marek Wodzinski, Andrzej Skalski, Izabela Ciepiela, Tomasz Kuszewski, Piotr
  Kedzierawski, and Janusz Gajda.
\newblock Improving oncoplastic breast tumor bed localization for radiotherapy
  planning using image registration algorithms.
\newblock {\em Physics in Medicine \& Biology}, 63(3):035024, 2018.

\bibitem{dromain2011dual}
Clarisse Dromain, Fabienne Thibault, Serge Muller, Fran{\c{c}}oise Rimareix,
  Suzette Delaloge, Anne Tardivon, and Corinne Balleyguier.
\newblock Dual-energy contrast-enhanced digital mammography: initial clinical
  results.
\newblock {\em European radiology}, 21(3):565--574, 2011.

\bibitem{iotti2017contrast}
Valentina Iotti, Sara Ravaioli, Rita Vacondio, Chiara Coriani, Sabrina
  Caffarri, Roberto Sghedoni, Andrea Nitrosi, Moira Ragazzi, Elisa Gasparini,
  Cristina Masini, et~al.
\newblock Contrast-enhanced spectral mammography in neoadjuvant chemotherapy
  monitoring: a comparison with breast magnetic resonance imaging.
\newblock {\em Breast Cancer Research}, 19(1):1--13, 2017.

\bibitem{barra2018contrast}
Filipe~Ramos Barra, Alaor~Barra Sobrinho, Renato~Ramos Barra, Mayra~Teixeira
  Magalh{\~a}es, Laira~Rodrigues Aguiar, Gabriela Feitosa Lins~de Albuquerque,
  Rodrigo~Pepe Costa, Luciano Farage, and Riccardo Pratesi.
\newblock Contrast-enhanced mammography ({CEM}) for detecting residual disease
  after neoadjuvant chemotherapy: a comparison with breast magnetic resonance
  imaging ({MRI}).
\newblock {\em BioMed research international}, 2018, 2018.

\bibitem{patel2018contrast}
Bhavika~K Patel, Talal Hilal, Matthew Covington, Nan Zhang, Heidi~E Kosiorek,
  Marc Lobbes, Donald~W Northfelt, and Barbara~A Pockaj.
\newblock Contrast-enhanced spectral mammography is comparable to {MRI} in the
  assessment of residual breast cancer following neoadjuvant systemic therapy.
\newblock {\em Annals of surgical oncology}, 25(5):1350--1356, 2018.

\bibitem{steinhof2021contrast}
Katarzyna Steinhof-Radwa{\'n}ska, Anna Gra{\.z}y{\'n}ska, Andrzej Lorek, Iwona
  Gisterek, Anna Barczyk-Gutowska, Agnieszka Bobola, Karolina Okas, Zuzanna
  Lelek, Irmina Morawska, Jakub Potoczny, et~al.
\newblock Contrast-enhanced spectral mammography assessment of patients treated
  with neoadjuvant chemotherapy for breast cancer.
\newblock {\em Current Oncology}, 28(5):3448--3462, 2021.

\bibitem{eisenhauer2009new}
Elizabeth~A Eisenhauer, Patrick Therasse, Jan Bogaerts, Lawrence~H Schwartz,
  Danielle Sargent, Robert Ford, Janet Dancey, S~Arbuck, Steve Gwyther,
  Margaret Mooney, et~al.
\newblock New response evaluation criteria in solid tumours: revised recist
  guideline (version 1.1).
\newblock {\em European journal of cancer}, 45(2):228--247, 2009.

\bibitem{moustafa2019quantitative}
Amr Farouk~Ibrahim Moustafa, Rasha~Mohammed Kamal, Mohammed Mohammed~Mohammed
  Gomaa, Shaimaa Mostafa, Roaa Mubarak, and Mohamed El-Adawy.
\newblock Quantitative mathematical objective evaluation of contrast-enhanced
  spectral mammogram in the assessment of response to neoadjuvant chemotherapy
  and prediction of residual disease in breast cancer.
\newblock {\em Egyptian Journal of Radiology and Nuclear Medicine},
  50(1):1--13, 2019.

\bibitem{kamal2020predicting}
Rasha~Mohammed Kamal, Sherihan~Mahmoud Saad, Amr Farouk~Ibrahim Moustafa,
  Mohammed~Mohammed Gomaa, Omniya Mokhtar, Iman Gouda, Ahmed Hassan, Amany
  Hilal, and Ashraf ElZayat.
\newblock Predicting response to neo-adjuvant chemotherapy and assessment of
  residual disease in breast cancer using contrast-enhanced spectral
  mammography: a combined qualitative and quantitative approach.
\newblock {\em Egyptian Journal of Radiology and Nuclear Medicine},
  51(1):1--14, 2020.

\bibitem{xing2021quantitative}
Dong Xing, Ning Mao, Jianjun Dong, Heng Ma, Qianqian Chen, and Yongbin Lv.
\newblock Quantitative analysis of contrast enhanced spectral mammography grey
  value for early prediction of pathological response of breast cancer to
  neoadjuvant chemotherapy.
\newblock {\em Scientific Reports}, 11(1):1--9, 2021.

\bibitem{wang2021contrast}
Zhongyi Wang, Fan Lin, Heng Ma, Yinghong Shi, Jianjun Dong, Ping Yang, Kun
  Zhang, Na~Guo, Ran Zhang, Jingjing Cui, et~al.
\newblock Contrast-enhanced spectral mammography-based radiomics nomogram for
  the prediction of neoadjuvant chemotherapy-insensitive breast cancers.
\newblock {\em Frontiers in Oncology}, 11:84, 2021.

\bibitem{van2003comparison}
Saskia van Engeland, Peter Snoeren, JHCL Hendriks, and Nico Karssemeijer.
\newblock A comparison of methods for mammogram registration.
\newblock {\em IEEE Transactions on Medical Imaging}, 22(11):1436--1444, 2003.

\bibitem{li2015bilateral}
Yanfeng Li, Houjin Chen, Yongyi Yang, Lin Cheng, and Lin Cao.
\newblock A bilateral analysis scheme for false positive reduction in mammogram
  mass detection.
\newblock {\em Computers in biology and medicine}, 57:84--95, 2015.

\bibitem{richard2003new}
Fr{\'e}d{\'e}ric~JP Richard and Laurent~D Cohen.
\newblock A new image registration technique with free boundary constraints:
  application to mammography.
\newblock {\em Computer Vision and Image Understanding}, 89(2-3):166--196,
  2003.

\bibitem{diez2011revisiting}
Yago D{\'\i}ez, Arnau Oliver, Xavier Llado, Jordi Freixenet, Joan Marti,
  Joan~Carles Vilanova, and Robert Marti.
\newblock Revisiting intensity-based image registration applied to mammography.
\newblock {\em IEEE Transactions on Information Technology in Biomedicine},
  15(5):716--725, 2011.

\bibitem{garcia2017similarity}
Eloy Garc{\'\i}a, Arnau Oliver, Yago Diez, Oliver Diaz, Xavier Llad{\'o},
  Robert Mart{\'\i}, and Joan Mart{\'\i}.
\newblock Similarity metrics for intensity-based registration using breast
  density maps.
\newblock In {\em Iberian Conference on Pattern Recognition and Image
  Analysis}, pages 217--225. Springer, 2017.

\bibitem{zhang2019mammographic}
Linlin Zhang, Yanfeng Li, Houjin Chen, and Lin Cheng.
\newblock Mammographic mass detection by bilateral analysis based on
  convolution neural network.
\newblock In {\em 2019 IEEE International Conference on Image Processing
  (ICIP)}, pages 784--788. IEEE, 2019.

\bibitem{haskins2020deep}
Grant Haskins, Uwe Kruger, and Pingkun Yan.
\newblock Deep learning in medical image registration: A survey.
\newblock {\em Machine Vision and Applications}, 31(1):8, 2020.

\bibitem{marti2014detecting}
Robert Mart{\'\i}, Yago D{\'\i}ez, Arnau Oliver, Meritxell Tortajada, Reyer
  Zwiggelaar, and Xavier Llad{\'o}.
\newblock Detecting abnormal mammographic cases in temporal studies using image
  registration features.
\newblock In {\em International Workshop on Digital Mammography}, pages
  612--619. Springer, 2014.

\bibitem{mendel2019temporal}
Kayla Mendel, Hui Li, Nabihah Tayob, Randa El-Zein, Isabelle Bedrosian, and
  Maryellen Giger.
\newblock Temporal mammographic registration for evaluation of architecture
  changes in cancer risk assessment.
\newblock In {\em Medical Imaging 2019: Computer-Aided Diagnosis}, volume
  10950, page 1095041. International Society for Optics and Photonics, 2019.

\bibitem{robinson2019machine}
Kayla~Rae Robinson.
\newblock {\em Machine Learning on Medical Imaging for Breast Cancer Risk
  Assessment}.
\newblock PhD thesis, The University of Chicago, 2019.

\bibitem{gennaro2022artifact}
Gisella Gennaro, Enrica Baldan, Elisabetta Bezzon, and Francesca Caumo.
\newblock Artifact reduction in contrast-enhanced mammography.
\newblock {\em Insights into Imaging}, 13(1):1--11, 2022.

\bibitem{zach2007duality}
Christopher Zach, Thomas Pock, and Horst Bischof.
\newblock A duality based approach for realtime tv-l 1 optical flow.
\newblock In {\em Joint pattern recognition symposium}, pages 214--223.
  Springer, 2007.

\bibitem{wedel2009improved}
Andreas Wedel, Thomas Pock, Christopher Zach, Horst Bischof, and Daniel
  Cremers.
\newblock An improved algorithm for tv-l 1 optical flow.
\newblock In {\em Statistical and geometrical approaches to visual motion
  analysis}, pages 23--45. Springer, 2009.

\bibitem{chambolle2011first}
Antonin Chambolle and Thomas Pock.
\newblock A first-order primal-dual algorithm for convex problems with
  applications to imaging.
\newblock {\em Journal of mathematical imaging and vision}, 40(1):120--145,
  2011.

\bibitem{sutton2009image}
Michael~A Sutton, Jean~Jose Orteu, and Hubert Schreier.
\newblock {\em Image correlation for shape, motion and deformation
  measurements: basic concepts, theory and applications}.
\newblock Springer Science \& Business Media, 2009.

\bibitem{Grediac}
M.~Gr\'ediac and F.~Hild, editors.
\newblock {\em {Full-field measurements and identification in solid
  mechanics}}.
\newblock John Wiley $\&$ Sons, 2012.

\bibitem{lee2019validation}
Lik~Chuan Lee and Martin Genet.
\newblock Validation of equilibrated warping—image registration with
  mechanical regularization—on 3d ultrasound images.
\newblock In {\em International Conference on Functional Imaging and Modeling
  of the Heart}, pages 334--341. Springer, 2019.

\bibitem{Besnard06}
G.~Besnard, F.~Hild, and S.~Roux.
\newblock ``finite-element'' displacement fields analysis from digital images:
  application to portevin--le ch{\^a}telier bands.
\newblock {\em Experimental Mechanics}, 46(6):789--803, 2006.

\bibitem{mendoza2019correlation}
Arturo Mendoza, Julien Schneider, Estelle Parra, and St{\'e}phane Roux.
\newblock The correlation framework: Bridging the gap between modeling and
  analysis for 3d woven composites.
\newblock {\em Composite Structures}, 229:111468, 2019.

\bibitem{krumm2008reducing}
Michael Krumm, Stefan Kasperl, and Matthias Franz.
\newblock Reducing non-linear artifacts of multi-material objects in industrial
  3d computed tomography.
\newblock {\em Ndt \& E International}, 41(4):242--251, 2008.

\bibitem{tikhonov1977solutions}
Andrey~N Tikhonov and Vasiliy~Y Arsenin.
\newblock Solutions of ill-posed problems.
\newblock {\em New York}, pages 1--30, 1977.

\bibitem{rethore2009extended}
Julien R{\'e}thor{\'e}, St{\'e}phane Roux, and Fran{\c{c}}ois Hild.
\newblock An extended and integrated digital image correlation technique
  applied to the analysis of fractured samples: The equilibrium gap method as a
  mechanical filter.
\newblock {\em European Journal of Computational Mechanics/Revue Europ{\'e}enne
  de M{\'e}canique Num{\'e}rique}, 18(3-4):285--306, 2009.

\bibitem{ostergaard2008acceleration}
Karsten {\O}stergaard~Noe, Baudouin~Denis De~Senneville, Ulrik~Vindelev
  Elstr{\o}m, Kari Tanderup, and Thomas~Sangild S{\o}rensen.
\newblock Acceleration and validation of optical flow based deformable
  registration for image-guided radiotherapy.
\newblock {\em Acta Oncologica}, 47(7):1286--1293, 2008.

\end{thebibliography}

\appendix

\section{Appendix}
\label{annexB}

\begin{figure}[ht!]
	\centering
        \includegraphics[width=0.5\textwidth]{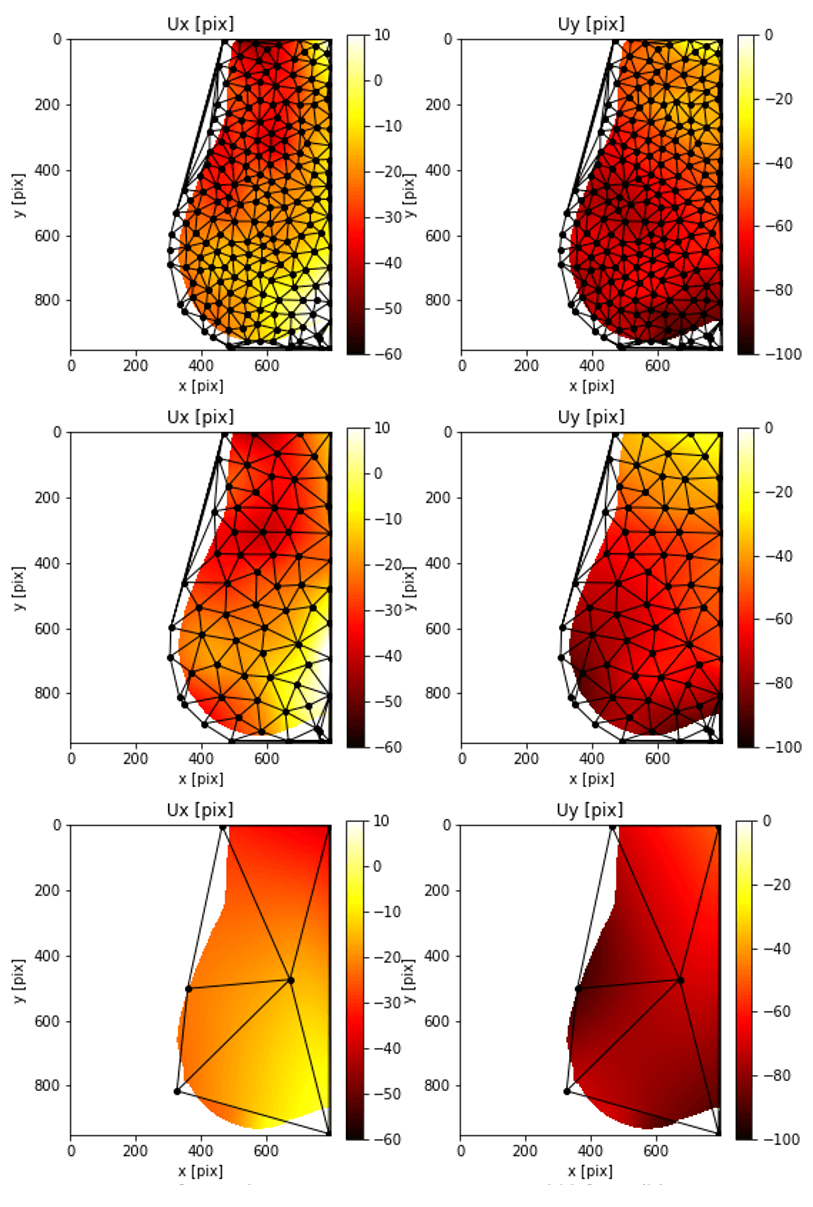}
	\caption{Motion results for three mesh sizes. The top, middle, and bottom results are respectively Fine, medium, and very coarse mesh. The results are here presented at scale 1/3 so 1~pix$\leftrightarrow$1/30~mm.}
	\label{fig:mesh_dep}  
\end{figure}

\paragraph{Discussion on the mesh dependency} To evaluate the impact of the kinematic support, different mesh sizes were used and the kinematic results are shown in Figure~\ref{fig:mesh_dep}. The metrics for the different mesh sizes are shown in Table~\ref{tab:mesh_kin}.

\begin{table}[ht!]
\centering
\begin{tabular}{c|c|c|c|}
\cline{2-4}
                                                              & Coarse mesh & Medium mesh & Fine mesh \\ \hline
\multicolumn{1}{|c|}{Kinematics mesh density {[}cm$^{-2}${]}} & 18          & 5           & 2         \\ \hline
\multicolumn{1}{|c|}{Intensity mesh density {[}cm$^{-2}${]}}  & 2           & 2           & 2         \\ \hline
\multicolumn{1}{|c|}{RMSE$[d]$ {[}mm{]}}                            & 10.2        & 5.9         & 5.8       \\ \hline
\multicolumn{1}{|c|}{Max$[d]$ {[}mm{]}}                   & 33.9        & 21.3         & 22.1      \\ \hline
\multicolumn{1}{|c|}{Nb of converged cases} & 207/208 (99.5\%) & 208/208 (100\%) & 208/208 (100\%) \\ \hline
\end{tabular}
\caption{Registration of clinical landmarks with different kinematics mesh sizes.}
\label{tab:mesh_kin}
\end{table}

A very coarse mesh is essentially based on the registration of the breast edges where the image gradient is very high. The registration metrics for the coarse mesh are hence not optimal.

When the mesh is very fine, the degrees of freedom are mainly damped by the regularization function. The problem became not mesh dependent as the metrics are not sensitive to the mesh evolution.

Finally, from the proposed metrics, the measured kinematics is not mesh-dependent and a choice is to select the mesh for fast computation.

\section{Appendix}
\label{annexA}
\paragraph{Intensity correction impacts on a real case}

\begin{figure}[ht!]
	\centering
	    \includegraphics[width=0.25\textwidth]{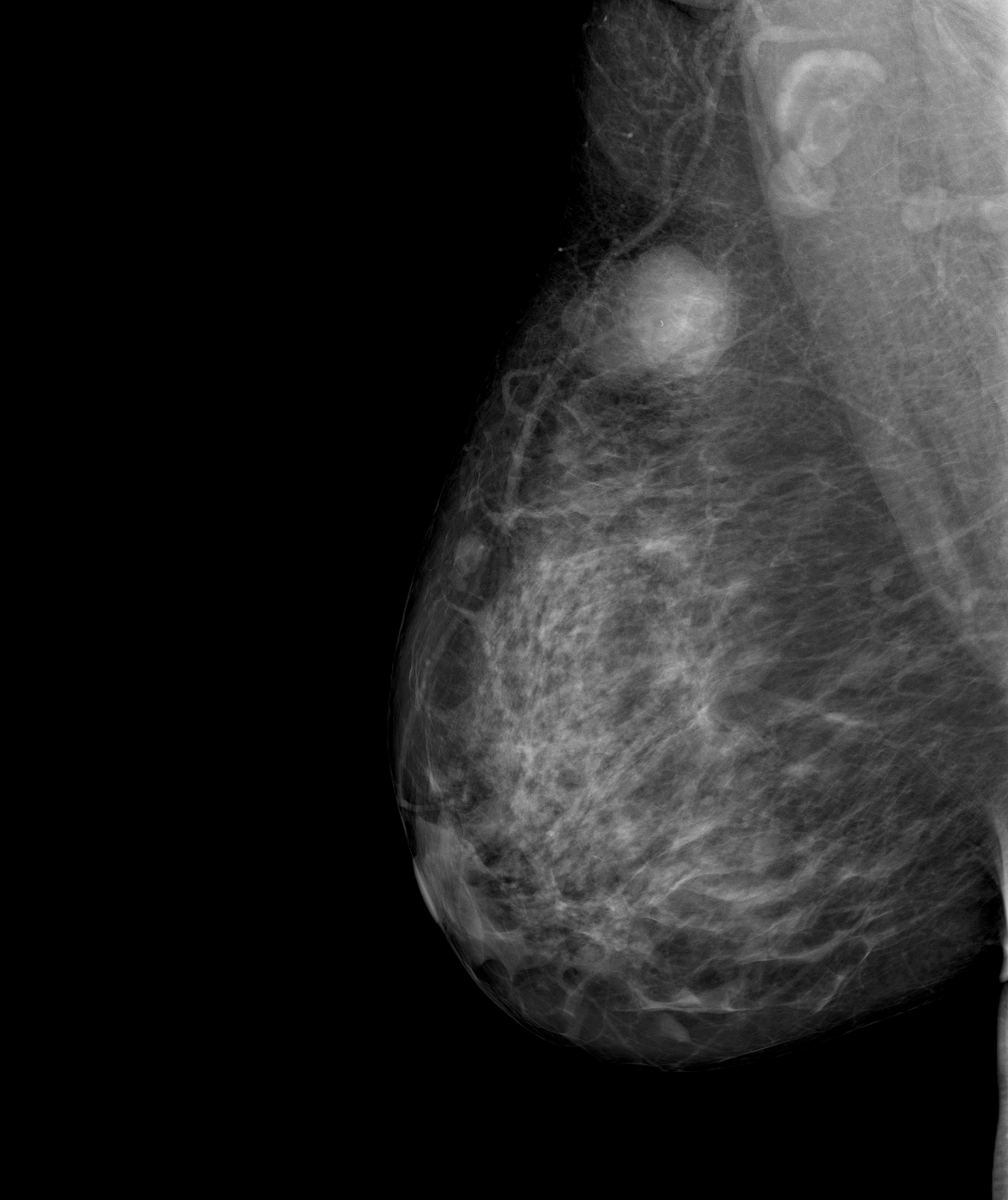}
        \includegraphics[width=0.25\textwidth]{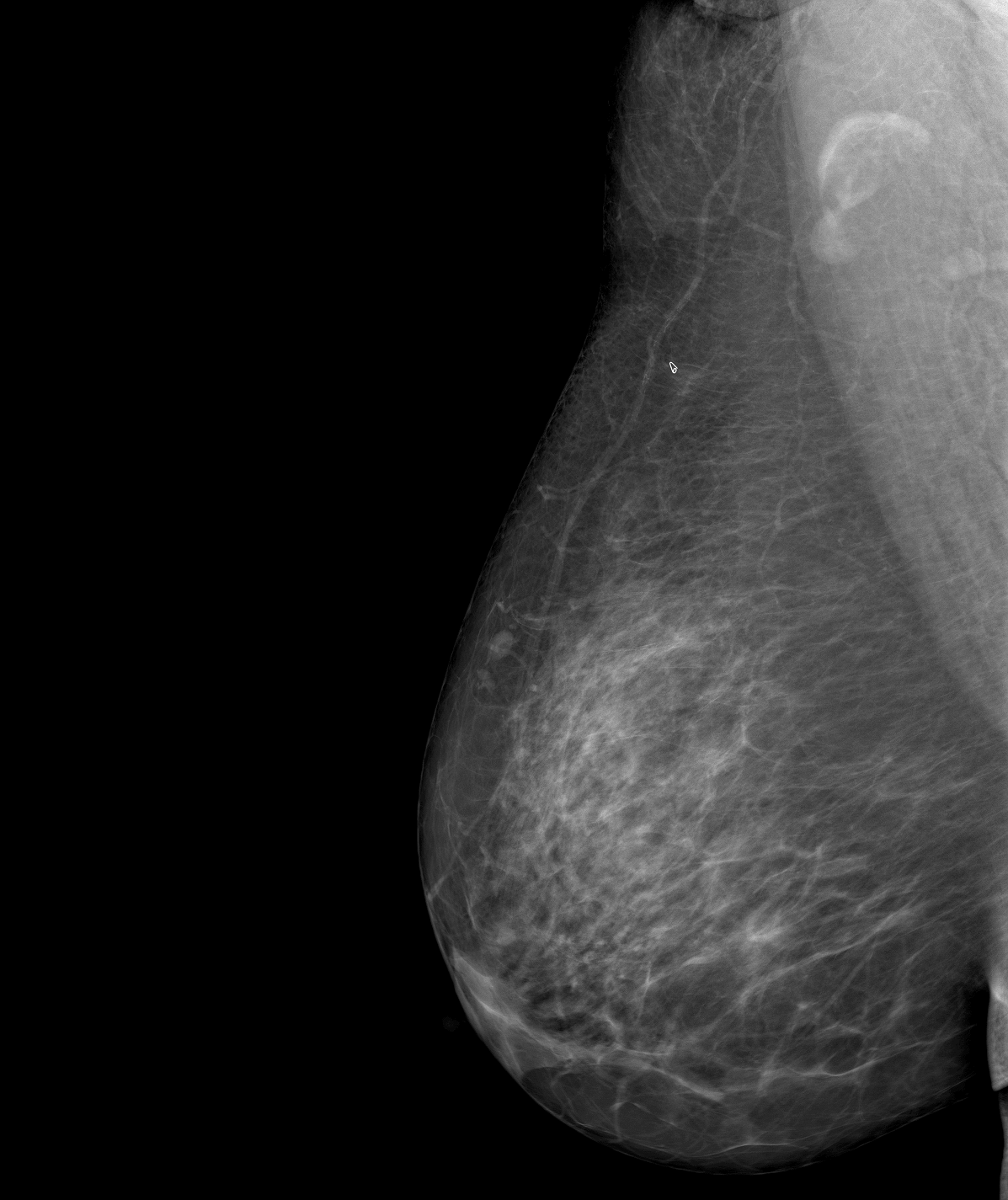}
        \includegraphics[width=0.6\textwidth]{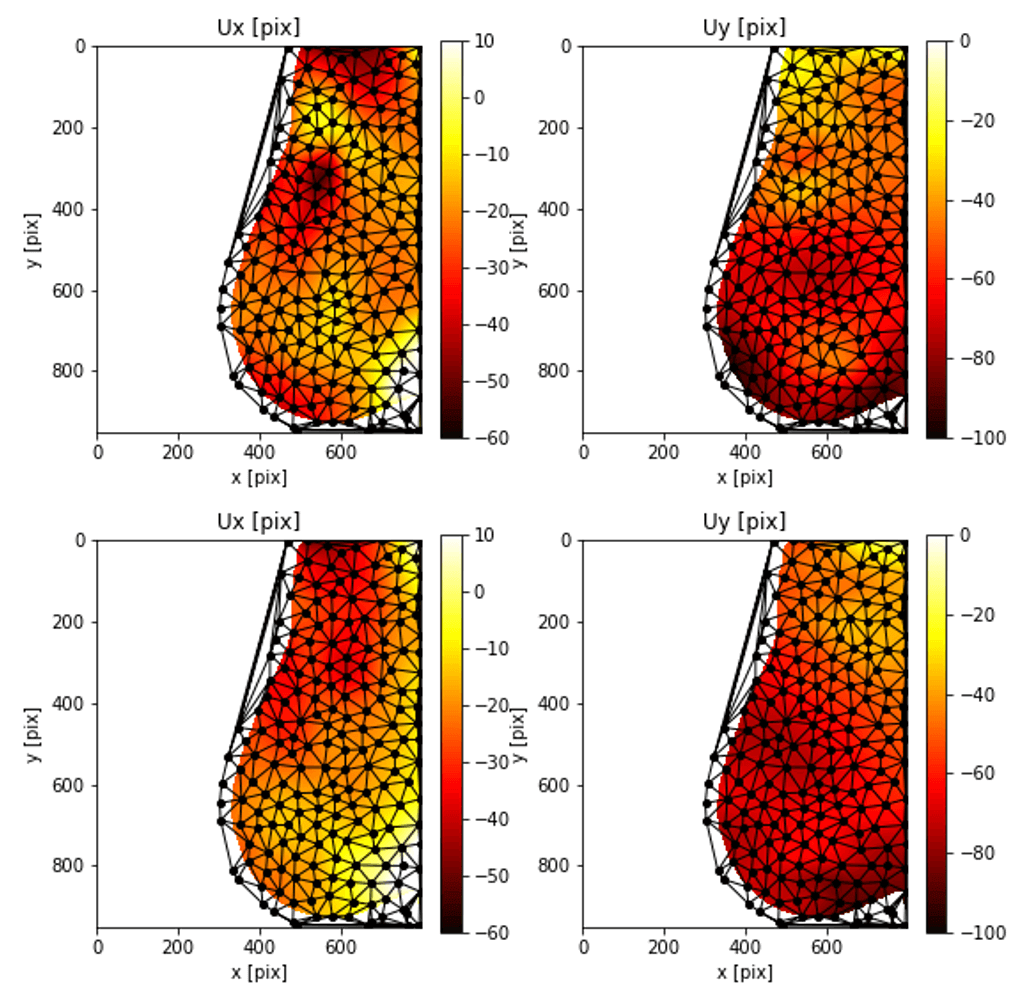}
	\caption{
	First line: a real clinical example with LE images pre-NAC (left) and post-NAC (right). 
	Second line: Displacement results with GDIC. This measurement is impacted by lesion evolution. 
	Third line: displacement with GDIC-I not impacted by the lesion evolution.}
	\label{fig:BC}  
\end{figure}

The local intensity correction allows correcting the lesion intensity evolution. As a result, the registration is not biased by the important texture evolution. The registration with and without intensity correction is shown in Figure~\ref{fig:BC}. Without a ground truth, it is not easy to evaluate the obtained motion. However, it can be seen that the measured motion aims at shrinking the lesion size (important positive and negative motion highlighting a compression field) to reduce its intensity impact. The GDIC-I motion is much smoother and more realistic. The intensity correction allows not to bias displacement measurement.

% ------------------------------------------------------------------------------
% ------------------------------------------------------------------------------
% ------------------------------------------------------------------------------
% ------------------------------------------------------------------------------
% ------------------------------------------------------------------------------
% ------------------------------------------------------------------------------
% ------------------------------------------------------------------------------
% ------------------------------------------------------------------------------
% ------------------------------------------------------------------------------
% ------------------------------------------------------------------------------
% ------------------------------------------------------------------------------
% ------------------------------------------------------------------------------

\end{document}